\documentclass[onecolumn,showpacs,preprintnumbers,10pt,superscriptaddress,nofootinbib]{revtex4}
\usepackage{amssymb}
\usepackage{amsmath}
\usepackage{graphicx}
\usepackage{dcolumn}
\usepackage{bm}

\input{tcilatex}

\begin{document}

\title{On the anomalous dimensions of the multiple pomeron exchanges}
\author{Bo-Wen Xiao}
\email{bowen@phys.columbia.edu}
\affiliation{Department of Physics, Columbia University, New York, NY, 10027, USA}

\begin{abstract}
High energy hard scattering in large $N_{c}$ limit can be described by the
QCD dipole model. In this paper, single, double and triple BFKL pomeron
exchange amplitudes are computed explicitly within the dipole model. Based
on the calculation, the general formula $\gamma^{(k)*}_{0}=\chi^{-1}\left(
k\chi\left(\frac{1}{2}\right)\right)$ which governs the anomalous dimension
of $1\Rightarrow k$ amplitude is conjectured. As far as the unitarity
problem is concerned, we find that the anomalous dimension $\gamma$ varies
from graph to graph due to the DGLAP evolution. In the end, a comparison
between this computation and reggeon field theory is provided.
\end{abstract}

\date{\today }
\maketitle

\section{Introduction}

High energy QCD small-x evolution of hadron can be described by the
Balitsky-Fadin-Kuraev-Lipatov (BFKL) Pomeron\cite%
{Balitsky:1978ic,Kuraev:1977fs,Lipatov:1985uk}. BFKL pomeron should be
relevant for describing the exponential growth of scattering amplitudes with
respect to rapidity. More than a decade ago, the dipole picture of pomeron
was discovered by Mueller et al\cite%
{Mueller:1993rr,Mueller:1994jq,Mueller:1994gb}, and the exact equivalence%
\cite{Navelet:1997tx} between the color dipole model and the BFKL pomeron
result was verified several years later. Onium-onium scattering at high
energy is used to describe this dipole picture of high energy hard
scattering in large $N_{c}$ limit. Supposing that $M$, the mass of the
onium, is large enough to justify the fixed coupling calculation, one can
find that the onium-onium scattering cross section at single pomeron
exchange level behaves as
\begin{equation}
\sigma \sim xx^{\prime }\alpha ^{2}\exp \left[ \left( \alpha _{P}-1\right) Y%
\right]
\end{equation}%
where $x$, $x^{\prime }$ are the sizes of the interacting onia, $Y=\ln \frac{%
s}{M^{2}}$ and $\alpha _{P}-1=\frac{8\alpha C_{F}\ln 2}{\pi }$ with $C_{F}=%
\frac{N_{c}^{2}-1}{2N_{c}}\Rightarrow \frac{N_{c}}{2}$ in large $N_{c}$
limit. Besides the exponential growth of the cross section, one also gets
the anomalous dimension $\gamma =\frac{1}{2}$ as a result of $\sigma $ being
proportional to $xx^{\prime }$.(The shift of the exponents of $x^{2}$ and $
x^{\prime 2}$ from $1$ is called the anomalous dimension of the BFKL
pomeron.)

On the other hand, single pomeron exchange amplitude violates the unitarity
and the Froissart bound at extremely high energy. Multiple exchanges of
pomerons (here we mean pomeron loops) should be able to reduce the growth
rate of the amplitudes, and eventually unitarize it at high energy. Within
the dipole model, Kovchegov equation\cite{Kovchegov:1999yj}, a nonlinear
evolution equation, is derived by re-summing the fan diagrams(i.e., multiple
pomeron exchanges in dipole-nucleus scattering). Recently, there has been a
lot of development\cite%
{Mueller:2004se,Iancu:2004es,Munier:2003vc,Iancu:2004iy,Mueller:2005ut,Iancu:2005nj,Kovner:2005nq,Hatta:2005rn}
in high-energy QCD evolution. Evolution equations which include pomeron
loops are derived\cite%
{Kovner:2005en+X,Blaizot:2005vf+X,Marquet:2005hu,Levin:2005au,Enberg:2005cb,Soyez:2005ha}
and utilized to QCD phenomenology(e.g., see \cite%
{Hatta:2006hs,Iancu:2006uc,Kozlov:2006qw,Kozlov:2007wm}.). As far as the
unitarity problem is concerned, one has to integrate over rapidity and
stretch the pomeron loops as large as possible in rapidity space in order to
obtain the maximum leading order loop amplitudes. Meanwhile, the anomalous
dimensions of pomeron loops are determined by the lower and upper pomerons
with infinitesimal rapidity length. Assuming that only one pomeron can
interact with the target or projectile onia, these two pomerons then connect
the target onium and projectile onium, respectively. In spite of having the
same characteristic function as other large rapidity pomeron, these two
pomeron, however, give rise to different value of anomalous dimensions to
the loop amplitudes\cite{Mueller:1994jq,Mueller:1994gb,Navelet:2002zz}.

As building blocks of pomeron loops, $1\Rightarrow 2$ amplitude as well as
triple pomeron vertex, which features dipole pair correlation in onium
states, has been studied extensively\cite%
{Mueller:1994jq,Peschanski:1997yx,Braun:1997nu,Hatta:2007fg,Korchemsky:1997fy,Navelet:2002zz,Levin:2007wc}%
. One can easily obtain the anomalous dimension of one loop amplitudes by
calculating the $1\Rightarrow 2$ amplitude $n_{Y}^{(2)}$ since the one loop
diagram is just nothing but square of $1\Rightarrow 2$ by hooking them in
the middle of the rapidity. Following the same reason, the $1\Rightarrow 3$
amplitude indicates the anomalous dimension of two loop amplitudes. The
objective of this paper is then to investigate the $1\Rightarrow 2$ and $%
1\Rightarrow 3$ amplitudes explicitly after integrating the rapidity from 0
to the maximum rapidity of the system $Y$, and then generalize the results
to $1\Rightarrow k$ amplitudes.

In this paper, in the way described above, we explicitly calculate the
anomalous dimensions of the single, double pomeron and triple pomeron
exchange amplitudes in QCD dipole model in the leading logarithmic
approximation. In addition to the usual pomeron anomalous dimension $\frac{1%
}{2}$, we find that $\gamma _{0}^{(2)\ast }=0.18$ and $\gamma _{0}^{(3)\ast
}=0.12$ for double and triple pomeron exchange amplitude, respectively. The
anomalous dimension $\gamma _{0}^{(2)\ast }=0.18$ of one loop amplitude is
not new, and it has been obtained in ref.~\cite%
{Mueller:1994jq,Mueller:1994gb,Navelet:2002zz}. The anomalous dimension of
the triple pomeron exchange amplitude $\gamma _{0}^{(3)\ast }=0.12$ is new
and is given by Eqs.~(\ref{n3f}) and ~(\ref{f3}). Based on the calculation,
the general formula $\gamma _{0}^{(k)\ast }=\chi ^{-1}\left( k\chi \left(
\frac{1}{2}\right) \right) $ which governs anomalous dimensions of $%
1\Rightarrow k$ amplitude is conjectured in the region $\ln ^{2}\left( q\rho
\right) \ll \frac{14\alpha C_{F}\zeta \left( 3\right) }{\pi }Y$ and $\rho
^{2}q^{2}\ll 1$, where $\rho $ stands for all dipole sizes and $q$
represents all momentum scales.

The paper is organized as follows: we start with the computation on single
pomeron exchange amplitude in onium-onium scattering and discussion on the
DGLAP evolution in double logarithmic limit, then calculate the double and
triple pomeron exchange amplitudes, which are then followed by the
generalization to $1\Rightarrow k$ amplitude and a comparison with the
reggeon field theory, as well as conclusions.

\section{The Dipole density in the QCD dipole model.}

\label{single}

\subsection{Single BFKL pomeron exchange in onium-onium scattering.}

In order to be intuitive and complete, let us first sketch the well-known
single pomeron exchange amplitude in onium-onium scattering. Following \cite%
{Mueller:1994jq,Navelet:1997tx,Peschanski:1997yx}, the distribution of
dipoles in an onium state in coordinate space reads,
\begin{eqnarray}
n_{Y}^{\left( 1\right) }\left( \rho _{0}\rho _{1};\rho _{a_{0}}\rho
_{a_{1}}\right) &=&\sum_{n=-\infty }^{\infty }\int d\nu \frac{2\nu
^{2}+n^{2}/2}{\pi ^{4}}\dint \frac{d\omega }{2\pi i}\frac{e^{\omega Y}}{%
\omega -\omega \left( n,\nu \right) } \\
&&\times \frac{1}{\rho _{a}^{2}}\dint d^{2}\rho _{\gamma }E^{h,\overline{h}%
}\left( \rho _{0\gamma },\rho _{1\gamma }\right) E^{h,\overline{h}\ast
}\left( \rho _{a_{0}\gamma },\rho _{a_{1}\gamma }\right)  \notag
\end{eqnarray}%
where all the coordinates $\rho $ are complex coordinates in the two
dimensional transverse space,
\begin{eqnarray}
\overrightarrow{\rho } &=&\left( \rho _{x},\rho _{y}\right) \\
\rho &=&\rho _{x}+i\rho _{y}\text{ and }\overline{\rho }=\rho _{x}-i\rho _{y}
\end{eqnarray}%
$E^{h,\overline{h}}\left( \rho _{0\gamma }\rho _{1\gamma }\right) $ is the
eigenfunction of the $SL\left( 2,C\right) $ group, and
\begin{eqnarray}
E^{h,\overline{h}}\left( \rho _{0\gamma }\rho _{1\gamma }\right) &=&\left(
-1\right) ^{n}\left( \frac{\rho _{01}}{\rho _{0\gamma }\rho _{1\gamma }}%
\right) ^{h}\left( \frac{\overline{\rho }_{01}}{\overline{\rho }_{0\gamma }%
\overline{\rho }_{1\gamma }}\right) ^{\overline{h}}, \\
E^{h,\overline{h}\ast }\left( \rho _{0\gamma }\rho _{1\gamma }\right)
&=&E^{1-h,1-\overline{h}}\left( \rho _{0\gamma }\rho _{1\gamma }\right) ,
\end{eqnarray}%
with $\rho _{a}=\rho _{a_{0}}-\rho _{a_{1}}$, $\rho _{10}=\rho _{0}-\rho
_{1} $ and
\begin{eqnarray}
h &=&\frac{1-n}{2}+i\nu  \notag \\
\overline{h} &=&\frac{1+n}{2}+i\nu .
\end{eqnarray}%
We define
\begin{equation}
\omega \left( n,\nu \right) =\frac{4\alpha C_{F}}{\pi }\left[ \psi \left(
1\right) -\frac{1}{2}\psi \left( \frac{1+\left\vert n\right\vert }{2}+i\nu
\right) -\frac{1}{2}\psi \left( \frac{1+\left\vert n\right\vert }{2}-i\nu
\right) \right] ,
\end{equation}%
where $\psi \left( x\right) $ is the digamma function. For fixed values of $%
\nu$, one can easily find that $\omega \left( 0,\nu \right)$ is always
larger than $\omega \left( n,\nu \right)$ with nonzero values of $n$. Thus,
as one can find in later discussions, the $n=0$ contribution corresponds to
the dominant pomeron trajectory with a positive intercept, and $n \neq 0$
parts correspond to sub-dominant reggeon trajectories with intercepts being
equal to or less than $0$. For convenience, we also define $\chi \left( \nu
,n\right) =\psi \left( 1\right) -\frac{1}{2}\psi \left( \frac{1+\left\vert
n\right\vert }{2}+i\nu \right) -\frac{1}{2}\psi \left( \frac{1+\left\vert
n\right\vert }{2}-i\nu \right) $, where $\chi \left( \nu ,n\right) $ is a
real function of $\nu $ which is analytic in the strip $-\frac{1+\left\vert
n\right\vert }{2}<\func{Im}\nu <\frac{1+\left\vert n\right\vert }{2}$. For
real value of $\nu $, $\chi \left( \nu ,n\right) $ is symmetrical with
respect to $\nu =0$, and it has a global maximum value at $\nu =0$.

According to the Fourier transform of $E^{h,\overline{h}}\left( \rho
_{0\gamma }\rho _{1\gamma }\right) $,
\begin{equation}
E^{h,\overline{h}}\left( \rho _{0\gamma }\rho _{1\gamma }\right) =E^{h,%
\overline{h}}\left( R+\frac{1}{2}\rho ,R-\frac{1}{2}\rho \right) =\frac{%
b_{n,\nu }\left\vert \rho \right\vert }{2\pi ^{2}}\int \frac{d^{2}q}{\left(
2\pi \right) ^{2}}e^{-iqR}E_{q}^{n,\nu }\left( \rho \right) ,
\end{equation}%
with $R=\frac{1}{2}\left( \rho _{0}+\rho _{1}\right) -\rho _{\gamma }$, $%
\rho =\rho _{0}-\rho _{1}$\footnote{%
Hereafter, we use $\rho$ as an abbreviation of above definition for $\rho
_{0}-\rho _{1}$.} and
\begin{equation}
b_{n,\nu }=\frac{\pi ^{3}2^{4i\nu }}{\left\vert n\right\vert /2-i\nu }\frac{%
\Gamma \left( \left\vert n\right\vert /2-i\nu +1/2\right) \Gamma \left(
\left\vert n\right\vert /2+i\nu \right) }{\Gamma \left( \left\vert
n\right\vert /2+i\nu +1/2\right) \Gamma \left( \left\vert n\right\vert
/2-i\nu \right) }.
\end{equation}%
One can get rid of the impact parameter dependence and define $n_{Y}^{(1)}$
in momentum space,%
\begin{eqnarray}
n_{Y,q,q_{a}}^{(1)\rho ,\rho _{a}} &=&\int
d^{2}rd^{2}r_{a}e^{iqr+q_{a}r_{a}}n_{Y}^{(1)}\left( \rho _{0}\rho _{1};\rho
_{a_{0}}\rho _{a_{1}}\right) \\
&=&\sum_{n=-\infty }^{\infty }\int d\nu \frac{2\nu ^{2}+n^{2}/2}{\pi ^{4}}%
\frac{1}{\rho _{a}^{2}}4\pi ^{2}\delta ^{\left( 2\right) }\left(
q-q_{a}\right)  \notag \\
&&\times \dint \frac{d\omega }{2\pi i}\frac{e^{\omega Y}}{\omega -\omega
\left( n,\nu \right) }\frac{b_{n,\nu }b_{n,\nu }^{\ast }\left\vert \rho \rho
_{a}\right\vert }{\left( 2\pi ^{2}\right) ^{2}}E_{q}^{n,\nu }\left( \rho
\right) E_{q_{a}}^{n,\nu \ast }\left( \rho _{a}\right) ,
\end{eqnarray}%
where $r=\frac{1}{2}\left( \rho _{0}+\rho _{1}\right) $ and $r_{a}=\frac{1}{2%
}\left( \rho _{a_{0}}+\rho _{a_{1}}\right) $. At very large rapidity $Y$,
the term corresponding to $n=0$ dominates the exponent since $\omega \left(
0,\nu \right) $ is greater than other $\omega \left( n,\nu \right) $. Thus,
all the contributions of sub-dominant trajectories (contributions from
nonzero $n$) can be neglected. Therefore, defining $n_{Y,q,q_{a}}^{(1)\rho
,\rho _{a}}=n_{Y,q}^{(1)\rho ,\rho _{a}}4\pi ^{2}\delta ^{\left( 2\right)
}\left( q-q_{a}\right) \,$, and noting that
\begin{equation}
\frac{b_{0,\nu }\left\vert \rho \right\vert }{2\pi ^{2}}E_{q}^{0,\nu }\left(
\rho \right) =\int d^{2}Re^{iqR}\left[ \frac{\rho }{\left\vert R-\frac{\rho
}{2}\right\vert \left\vert R+\frac{\rho }{2}\right\vert }\right] ^{1+2i\nu },
\end{equation}%
one gets
\begin{eqnarray}
n_{Y,q}^{(1)\rho ,\rho _{a}} &\simeq &\int d\nu \frac{2\nu ^{2}}{\pi ^{4}}%
\frac{1}{\rho _{a}^{2}}\frac{b_{0,\nu }b_{0,\nu }^{\ast }\left\vert \rho
\rho _{a}\right\vert }{\left( 2\pi ^{2}\right) ^{2}}E_{q}^{0,\nu }\left(
\rho \right) E_{q}^{0,\nu \ast }\left( \rho _{a}\right) \exp \left[ \omega
\left( 0,\nu \right) Y\right] ,  \label{n1q} \\
&\simeq &\frac{1}{2}\frac{\left\vert q\rho \right\vert }{\left\vert q\rho
_{a}\right\vert }e\left( \rho ,q\right) e\left( \rho _{a},-q\right) \frac{%
\exp \left[ \left( \alpha _{P}-1\right) Y\right] \exp \left[ -\frac{\pi \ln
^{2}\left( \rho /\rho _{a}\right) }{28\alpha C_{F}\zeta \left( 3\right) Y}%
\right] }{\left( 7\alpha \zeta \left( 3\right) C_{F}Y\right) ^{3/2}},
\end{eqnarray}%
with the famous BFKL pomeron intercept $\alpha _{P}-1=\frac{8\alpha C_{F}\ln
2}{\pi }$, and $q\rho $ and $q\rho _{a}$ being much smaller than $1$. In
arriving at the above result, we have used saddle point approximation and
assumed that diffusion approximation $\ln ^{2}\left( \rho /\rho _{a}\right)
\ll \frac{14\alpha C_{F}\zeta \left( 3\right) }{\pi }Y$ is valid, and have
defined
\begin{equation}
e\left( \rho _{a},q_{a}\right) =\frac{1}{2\pi }\int d^{2}R\exp \left(
iq_{a}R\right) \frac{1}{\left\vert R-\frac{\rho _{a}}{2}\right\vert
\left\vert R+\frac{\rho _{a}}{2}\right\vert }.
\end{equation}%
It is straightforward to see that the saddle point approximation picks the
dominant contribution of the $\int d\nu $ integral in the vicinity of $\nu
=0 $ which gives rise to the anomalous dimension of single BFKL pomeron: $%
\gamma ^{\ast }=\frac{1}{2}+i\nu ^{\ast }=\frac{1}{2}$.

Furthermore, employing saddle point approximation, it is easy to find a
general formula for the intercepts of all reggeon trajectories(for all
values of $n$): $\alpha_{n}-1=\frac{4\alpha C_{F}}{\pi }\chi \left(0
,n\right)$. For reader's convenience, we list the first three sets of
intercepts in the following: $\alpha_{0}-1=2.77 \frac{2\alpha C_{F}}{\pi }$,
$\alpha_{\pm 1 }-1=0$ and $\alpha_{\pm 2}-1=-1.23 \frac{2\alpha C_{F}}{\pi }$%
. The $n=0$ contribution gives to the dominant pomeron trajectory with a
positive intercept, and $n \neq 0$ parts correspond to sub-dominant reggeon
trajectories with intercepts being less than or equal to $0$.

In addition, the single pomeron exchange amplitude between two onia with
sizes $\rho $ and $\rho ^{\prime }$ then scales as
\begin{equation}
F^{\left( 1\right) }\left( \rho ,\rho ^{\prime },q,Y\right) \propto \left(
q^{2}\rho ^{2}\right) ^{\frac{1}{2}}\left( q^{2}\rho ^{\prime 2}\right) ^{%
\frac{1}{2}}\left\{ \alpha ^{2}\frac{\exp \left[ \left( \alpha _{P}-1\right)
Y\right] }{\left( 7\alpha \zeta \left( 3\right) C_{F}Y\right) ^{3/2}}%
\right\} .
\end{equation}

\subsection{The DGLAP evolution.}

In comparison, we would like to discuss another interesting limit of the one
dipole amplitude. Other than the extremely large $Y$ rapidity limit, we now
focus on the limit where $\ln \left( \frac{\rho _{a}}{\rho }\right) ^{2}\gg
\frac{4\alpha C_{F}}{\pi }Y$ and $\ln \frac{1}{q\rho }\gg 1$ while $Y$ is
relatively small. This calculation is useful to understand the new anomalous
dimensions of dipole pair and triplet densities found in later discussions.
Therefore, one can cast the one dipole amplitude (Eq.(\ref{n1q})) into the
form,
\begin{eqnarray}
n_{Y,q}^{(1)\rho ,\rho _{a}} &\simeq &\int d\nu \frac{1}{2\pi ^{2}}\frac{%
\left\vert \rho \rho _{a}\right\vert }{\rho _{a}^{2}}E_{q}^{0,\nu }\left(
\rho \right) E_{q}^{0,\nu \ast }\left( \rho _{a}\right) \exp \left[ \omega
\left( 0,\nu \right) Y\right] , \\
&\simeq &\int d\nu \frac{1}{2\pi ^{2}}\left[ \left( \frac{\rho ^{2}}{\rho
_{a}^{2}}\right) ^{\gamma }+\left( \frac{\rho ^{2}}{\rho _{a}^{2}}\right)
^{1-\gamma }\right] \exp \left[ \omega \left( 0,\nu \right) Y\right] , \\
&\simeq &\frac{1}{\pi ^{2}}\frac{\rho ^{2}}{\rho _{a}^{2}}\sqrt{\frac{\pi
\sqrt{\frac{4\alpha C_{F}}{\pi }Y}}{2\ln \left( \frac{\rho _{a}^{2}}{\rho
^{2}}\right) \sqrt{\ln \left( \frac{\rho _{a}^{2}}{\rho ^{2}}\right) }}}\exp %
\left[ 2\sqrt{\frac{4\alpha C_{F}}{\pi }Y\ln \left( \frac{\rho _{a}^{2}}{%
\rho ^{2}}\right) }\right]
\end{eqnarray}%
where one has used $b_{0,\nu }b_{0,\nu }^{\ast }=\frac{\pi ^{6}}{\nu ^{2}}$
and the expansion \cite{Navelet:1997tx,Navelet:1997xn}
\begin{equation}
\left. E_{q}^{0,\nu }\left( \rho \right) E_{q}^{0,\nu \ast }\left( \rho
_{a}\right) \right\vert _{q\ll \frac{1}{\rho },\frac{1}{\rho _{a}}}\simeq
\left( \frac{\rho ^{2}}{\rho _{a}^{2}}\right) ^{i\nu }+\left( \frac{\rho ^{2}%
}{\rho _{a}^{2}}\right) ^{-i\nu },
\end{equation}%
as well as saddle point approximation in the vicinity of $\gamma \simeq 1$
or $0$. Thus the cross section scales as,
\begin{equation}
\sigma \left( \rho ,\rho _{a},Y\right) \sim \alpha ^{2}\rho ^{2}\exp \left[ 2%
\sqrt{\frac{4\alpha C_{F}}{\pi }Y\ln \left( \frac{\rho _{a}^{2}}{\rho ^{2}}%
\right) }\right] .
\end{equation}%
In this collinear limit ($\rho _{a}\gg \rho $), we retrieve the well-known
result of the DGLAP evolution\cite{Gribov:1972ri} when the strong coupling
is fixed. The DGLAP evolution is featured by the exponent being $2\sqrt{%
\frac{4\alpha C_{F}}{\pi }Y\ln \left( \frac{\rho _{a}^{2}}{\rho ^{2}}\right)
}$ since it involves evolutions both in rapidity and virtuality (the
reciprocal of the dipole sizes) in the so-called double leading logarithmic
limit. Furthermore, as a result of being proportional to $\rho ^{2}$ when
the logarithmic dependence of $\rho ^{2}$ is neglected, the anomalous
dimension in this situation is zero, as opposed to the anomalous dimension
of the BFKL pomeron being $\frac{1}{2}$(for a pedagogical introduction of
BFKL and DGLAP evolutions, see \cite{Salam:1999cn}). In addition, in small $%
Y $ limit, this coincides with the perturbative QCD calculation of the
dipole-dipole cross section\cite{Navelet:1997tx,Mueller:1999yb}
\begin{equation}
\sigma _{dd}\left( x,x^{\prime }\right) =2\pi \alpha ^{2}x_{<}^{2}\left(
1+\ln \frac{x_{>}}{x_{<}}\right) ,
\end{equation}%
with $x_{>}$ being the greater one of $x$ and $x^{\prime }$ and $x_{<}$
being the lesser. Therefore, one expects that the anomalous dimension should
approach zero when rapidity is small or DGLAP limit (here we mean double
leading logarithmic limit which includes both evolutions in virtuality and
rapidity) is valid.

\section{The $1\Rightarrow 2$ amplitude and double pomeron exchange.}

\label{double2}

\begin{figure}[tbp]
\begin{center}
\includegraphics[width=4cm]{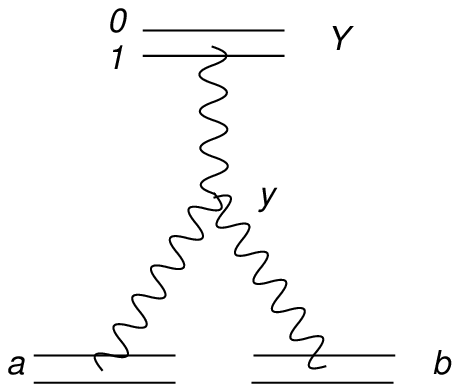} \includegraphics[width=5cm]{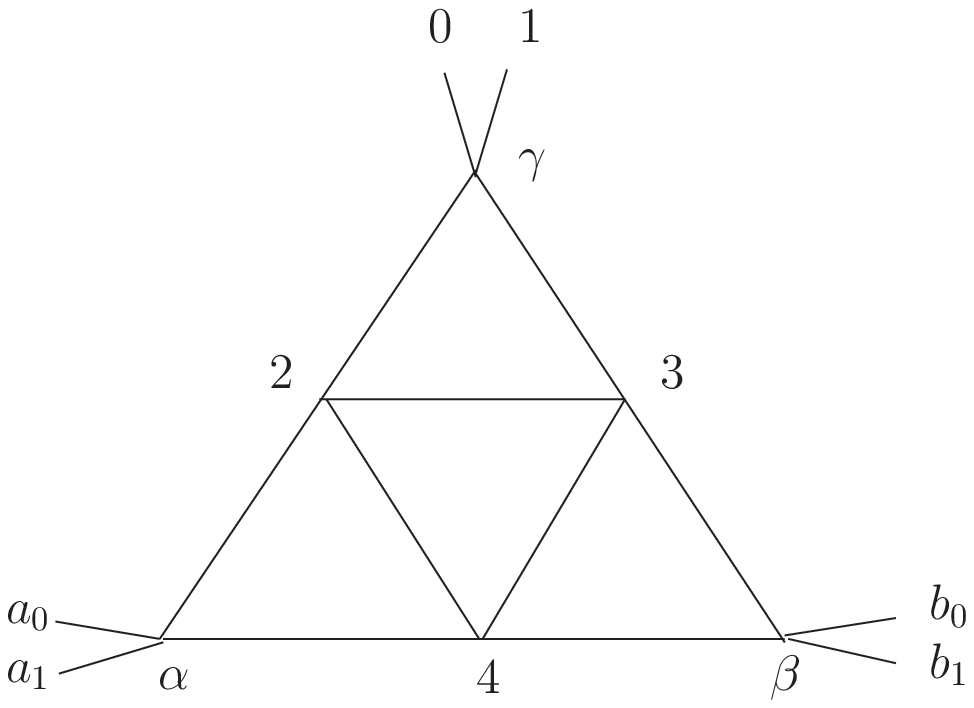}
\end{center}
\caption[*]{Dipole pair density $n^{2}_{Y}$ in an onium state with initial
size $\protect\rho =\protect\rho _{0}-\protect\rho _{1}$ and two children
dipoles having transverse separation $\protect\rho _{a}$ and $\protect\rho %
_{b}$. The left graph shows the rapidity structure of $n_{2} $, and the
right graphs indicates the graphical representation of the triple pomeron
vertex. }
\label{n2graph}
\end{figure}

As shown in Fig.~\ref{n2graph}, the $1\Rightarrow 2$ amplitude is defined as
the dipole pair density in an onium state. It involves a triple pomeron
vertex which splits the upper pomeron into two descendent pomerons. In QCD
dipole model\cite{Mueller:1993rr,Mueller:1994jq,Mueller:1994gb}, following
Peschanski\cite{Peschanski:1997yx}, one can write the $1\Rightarrow 2$
amplitude as
\begin{eqnarray}
&&n_{Y}^{(2)}\left( \rho _{0}\rho _{1};\rho _{a_{0}}\rho _{a_{1}},\rho
_{b_{0}}\rho _{b_{1}}\right)  \notag \\
&=&\dint dhdh_{a}dh_{b}\frac{1}{\rho _{a}^{2}\rho _{b}^{2}}\dint \frac{%
d\omega }{2\pi i}\frac{e^{\omega Y}}{\omega -\omega \left( n,\nu \right) }%
\frac{\frac{\alpha N_{c}}{\pi }}{\omega -\omega \left( a\right) -\omega
\left( b\right) }  \notag \\
&&\times \dint d^{2}\rho _{\alpha }d^{2}\rho _{\beta }d^{2}\rho _{\gamma
}E^{h,\overline{h}}\left( \rho _{0\gamma },\rho _{1\gamma }\right) E^{h_{a},%
\overline{h}_{b}}\left( \rho _{a_{0}\alpha },\rho _{a_{1}\alpha }\right)
E^{h_{b},\overline{h}_{b}}\left( \rho _{b_{0}\beta },\rho _{b_{1}\beta
}\right)  \notag \\
&&\times \dint \frac{d^{2}\rho _{2}d^{2}\rho _{3}d^{2}\rho _{4}}{\left\vert
\rho _{23}\rho _{34}\rho _{42}\right\vert ^{2}}E^{h,\overline{h}\ast }\left(
\rho _{2\gamma },\rho _{3\gamma }\right) E^{h_{a},\overline{h}_{b}\ast
}\left( \rho _{2\alpha },\rho _{4\alpha }\right) \overline{E}^{h_{b},%
\overline{h}_{b}\ast }\left( \rho _{3\beta },\rho _{4\beta }\right) , \\
&=&\dint dhdh_{a}dh_{b}\frac{1}{\rho _{a}^{2}\rho _{b}^{2}}\int_{0}^{Y}\frac{%
\alpha N_{c}}{\pi }dy\exp \left[ \omega \left( n,\nu \right) \left(
Y-y\right) +\left( \omega \left( a\right) +\omega \left( b\right) \right) y%
\right]  \notag \\
&&\times \dint d^{2}\rho _{\alpha }d^{2}\rho _{\beta }d^{2}\rho _{\gamma
}E^{h,\overline{h}}\left( \rho _{0\gamma },\rho _{1\gamma }\right) E^{h_{a},%
\overline{h}_{b}}\left( \rho _{a_{0}\alpha },\rho _{a_{1}\alpha }\right)
E^{h_{b},\overline{h}_{b}}\left( \rho _{b_{0}\beta },\rho _{b_{1}\beta
}\right)  \notag \\
&&\times \dint \frac{d^{2}\rho _{2}d^{2}\rho _{3}d^{2}\rho _{4}}{\left\vert
\rho _{23}\rho _{34}\rho _{42}\right\vert ^{2}}E^{h,\overline{h}\ast }\left(
\rho _{2\gamma },\rho _{3\gamma }\right) E^{h_{a},\overline{h}_{b}\ast
}\left( \rho _{2\alpha },\rho _{4\alpha }\right) \overline{E}^{h_{b},%
\overline{h}_{b}\ast }\left( \rho _{3\beta },\rho _{4\beta }\right)
\end{eqnarray}%
with $\rho _{a}=\rho _{a_{0}}-\rho _{a_{1}}$, $\rho _{b}=\rho _{b_{0}}-\rho
_{b_{1}}$ and
\begin{eqnarray}
\omega \left( a\right) &=&\frac{4\alpha C_{F}}{\pi }\left[ \psi \left(
1\right) -\frac{1}{2}\psi \left( \frac{1+\left\vert n_{a}\right\vert }{2}%
+i\nu _{a}\right) -\frac{1}{2}\psi \left( \frac{1+\left\vert
n_{a}\right\vert }{2}-i\nu _{a}\right) \right] ,  \notag \\
\omega \left( b\right) &=&\frac{4\alpha C_{F}}{\pi }\left[ \psi \left(
1\right) -\frac{1}{2}\psi \left( \frac{1+\left\vert n_{b}\right\vert }{2}%
+i\nu _{b}\right) -\frac{1}{2}\psi \left( \frac{1+\left\vert
n_{b}\right\vert }{2}-i\nu _{b}\right) \right] .
\end{eqnarray}%
We also use a compact notation,%
\begin{equation}
\int dh=\sum_{n=-\infty }^{\infty }\int d\nu \frac{2\nu ^{2}+n^{2}/2}{\pi
^{4}}.
\end{equation}%
Essentially, the above expression for $n_{Y}^{(2)}$ is the same as the ones
in ref.~\cite{Braun:1997nu,Hatta:2007fg}. On the other hand, hereafter, we
employ different methods of evaluation and show that there exists a
different anomalous dimension in $n_{Y}^{(2)}$ other than the typical
pomeron anomalous dimension $\gamma =1/2$.

First of all, the last line of the above definition of $n_{Y}^{(2)}$ can be
defined and written as%
\begin{eqnarray}
\overline{R}_{\gamma ,\alpha ,\beta }^{h,h_{a},h_{b}} &=&R_{\gamma ,\alpha
,\beta }^{1-h,1-h_{a},1-h_{b}},  \notag \\
&=&\dint \frac{d^{2}\rho _{2}d^{2}\rho _{3}d^{2}\rho _{4}}{\left\vert \rho
_{23}\rho _{34}\rho _{42}\right\vert ^{2}}E^{h,\overline{h}\ast }\left( \rho
_{2\gamma }\rho _{3\gamma }\right) E^{h_{a},\overline{h}_{b}\ast }\left(
\rho _{2\alpha }\rho _{4\alpha }\right) \overline{E}^{h_{b},\overline{h}%
_{b}\ast }\left( \rho _{3\beta }\rho _{4\beta }\right) .
\end{eqnarray}%
After using the $SL\left( 2,C\right) $ transformation
\begin{equation}
\rho ^{\prime }=\frac{\rho _{\beta }-\rho _{\gamma }}{\rho _{\beta }-\rho
_{\alpha }}\frac{\rho -\rho _{\alpha }}{\rho -\rho _{\gamma }},
\end{equation}%
one can easily write
\begin{equation}
R_{\gamma ,\alpha ,\beta }^{1-h,1-h_{a},1-h_{b}}=g_{3P}\left(
1-h,1-h_{a},1-h_{b}\right) \left( \rho _{\alpha \beta }^{-1+i}\rho _{\beta
\gamma }^{-1+j}\rho _{\gamma \alpha }^{-1+k}\right) \left( \overline{\rho }%
_{\alpha \beta }^{-1+\overline{i}}\overline{\rho }_{\beta \gamma }^{-1+%
\overline{j}}\overline{\rho }_{\gamma \alpha }^{-1+\overline{k}}\right) ,
\end{equation}%
where%
\begin{eqnarray}
i &=&-h+h_{a}+h_{b}, \\
j &=&-h_{a}+h_{b}+h, \\
k &=&-h_{b}+h_{a}+h,
\end{eqnarray}
and
\begin{eqnarray}
g_{3P}\left( h,h_{a},h_{b}\right) &=&\dint \frac{d^{2}\rho _{2}d^{2}\rho
_{3}d^{2}\rho _{4}}{\left\vert \rho _{23}\rho _{34}\rho _{42}\right\vert ^{2}%
}\rho _{23}^{h}\overline{\rho }_{23}^{\overline{h}}\left( \frac{\rho _{24}}{%
\rho _{2}\rho _{4}}\right) ^{h_{a}}\left( \frac{\overline{\rho }_{24}}{%
\overline{\rho }_{2}\overline{\rho }_{4}}\right) ^{\overline{h}_{a}}  \notag
\\
&&\times \left( \frac{\rho _{34}}{\left( 1-\rho _{3}\right) \left( 1-\rho
_{4}\right) }\right) ^{h_{b}}\left( \frac{\overline{\rho }_{34}}{\left( 1-%
\overline{\rho }_{3}\right) \left( 1-\overline{\rho }_{4}\right) }\right) ^{%
\overline{h}_{b}}
\end{eqnarray}%
Explicit evaluation of $g_{3P}$ can be found in ref.~(\cite%
{Korchemsky:1997fy}) and Appendix \ref{g3p}.

Furthermore, according to the Fourier transform of $E^{h,\overline{h}}\left(
\rho _{0\gamma }\rho _{1\gamma }\right) $, one can get rid of the impact
parameter dependence and define $n_{Y}^{(2)}$ in momentum space,
\begin{eqnarray}
n_{Y,Q,q_{a},q_{b}}^{(2)\rho ,\rho _{a},\rho _{b}} &=&\int
d^{2}rd^{2}r_{a}d^{2}r_{b}e^{iQr+q_{a}r_{a}+iq_{b}r_{b}}n_{Y}^{(2)}\left(
\rho _{0}\rho _{1};\rho _{a_{0}}\rho _{a_{1}},\rho _{b_{0}}\rho
_{b_{1}}\right)  \notag \\
&=&\dint dhdh_{a}dh_{b}\frac{1}{\rho _{a}^{2}\rho _{b}^{2}}\dint \frac{%
d\omega }{2\pi i}\frac{e^{\omega Y}}{\omega -\omega \left( n,\nu \right) }%
\frac{\frac{\alpha N_{c}}{\pi }}{\omega -\omega \left( a\right) -\omega
\left( b\right) }  \notag \\
&&\times \frac{b_{n,\nu }b_{n_{a},\nu _{a}}b_{n_{b},\nu _{b}}\left\vert \rho
\rho _{a}\rho _{b}\right\vert }{\left( 2\pi ^{2}\right) ^{3}}E_{Q}^{n,\nu
}\left( \rho \right) E_{q_{a}}^{n_{a},\nu _{a}}\left( \rho _{a}\right)
E_{q_{b}}^{n_{b},\nu _{b}}\left( \rho _{b}\right) g_{3P}\left(
1-h,1-h_{a},1-h_{b}\right)  \notag \\
&&\times \dint d^{2}\rho _{\alpha }d^{2}\rho _{\beta }d^{2}\rho _{\gamma
}\exp \left( iQ\rho _{\gamma }+iq_{a}\rho _{\alpha }+iq_{b}\rho _{\beta
}\right) \left[ \rho _{\alpha \beta }^{-1+i}\rho _{\beta \gamma }^{-1+j}\rho
_{\gamma \alpha }^{-1+k}\right] \left[ a.h.\right]
\end{eqnarray}%
where a.h. stands for the antiholomorphic part of the square-bracket term, $%
r=\frac{1}{2}\left( \rho _{0}+\rho _{1}\right) $, $r_{a}=\frac{1}{2}\left(
\rho _{a_{0}}+\rho _{a_{1}}\right) $ and $r_{b}=\frac{1}{2}\left( \rho
_{b_{0}}+\rho _{b_{1}}\right) $. Hereafter in this section, we compute $%
n_{Y}^{(2)}$ in $\rho ^{2}q_{a}^{2}\ll 1$ limit, where $q_{a}^{2}$, $%
q_{b}^{2}$ and $Q^{2}$ are of the same order and $\rho $, $\rho _{a}$ and $%
\rho _{b}$ are of the same order as well.

In addition, let us define
\begin{equation}
I_{Q,q_{a},q_{b}}^{h,h_{a},h_{b}}=\dint d^{2}\rho _{\alpha }d^{2}\rho
_{\beta }d^{2}\rho _{\gamma }\exp \left( iQ\rho _{\gamma }+iq_{a}\rho
_{\alpha }+iq_{b}\rho _{\beta }\right) \left[ \rho _{\alpha \beta
}^{-1+i}\rho _{\beta \gamma }^{-1+j}\rho _{\gamma \alpha }^{-1+k}\right] %
\left[ a.h.\right]
\end{equation}%
In order to evaluate $I_{Q,q_{a},q_{b}}^{h,h_{a},h_{b}}$, we change the
variables $\rho _{\alpha }$, $\rho _{\beta }$ and $\rho _{\gamma }$ into $u$%
, $v$ and $w$, where
\begin{eqnarray}
u &=&\frac{1}{2}\left( \rho _{\alpha }+\rho _{\beta }\right) ,  \notag \\
v &=&\rho _{\alpha }-\rho _{\beta },  \notag \\
w &=&\rho _{\alpha }-\rho _{\gamma }+\rho _{\beta }-\rho _{\gamma }.
\end{eqnarray}%
Then the integral can be cast into,
\begin{eqnarray}
I_{Q,q_{a},q_{b}}^{h,h_{a},h_{b}} &=&\frac{1}{4}\dint d^{2}ud^{2}vd^{2}w\exp
\left( i\left( Q+q_{a}+q_{b}\right) u-iQ\frac{w}{2}+i\left(
q_{a}-q_{b}\right) \frac{v}{2}\right)  \notag \\
&&\times \left[ v^{-1+i}\left( \frac{w-v}{2}\right) ^{-1+j}\left( \frac{w+v}{%
2}\right) ^{-1+k}\right] \left[ a.h.\right] ,  \notag \\
&=&\pi ^{2}\delta ^{\left( 2\right) }\left( Q+q_{a}+q_{b}\right) \dint
d^{2}vd^{2}w\exp \left( -iQ\cdot \frac{w}{2}+i\left( q_{a}-q_{b}\right)
\cdot \frac{v}{2}\right)  \notag \\
&&\times \left[ v^{-1+i}\left( \frac{w-v}{2}\right) ^{-1+j}\left( \frac{w+v}{%
2}\right) ^{-1+k}\right] \left[ a.h.\right] ,
\end{eqnarray}%
where the factor of $\frac{1}{4}$ comes from the Jacobian. Hereafter, in
order to simplify the calculation, we define
\begin{equation}
I_{Q,q_{a},q_{b}}^{h,h_{a},h_{b}}=\pi ^{2}\delta ^{\left( 2\right) }\left(
Q+q_{a}+q_{b}\right) I_{Q,q_{-}}^{h,h_{a},h_{b}},
\end{equation}%
with $q_{-}=\frac{1}{2}\left( q_{a}-q_{b}\right) $, perform the Taylor
expansion of the term $\exp \left( -iQ\frac{w}{2}\right) $ and only keep the
first term. Keeping only the first term is equivalent to the physical case
when $Q=0$ and $q_{a}=-q_{b}=q$ which corresponds to the forward scattering
of an onium on two nucleons. We put the evaluation of $%
I_{Q,q_{-}}^{h,h_{a},h_{b}}$ and discussion of higher order terms in the
expansion in Appendix \ref{integral} and \ref{general}, respectively. As
shown in the appendix, the following conclusion holds for all the terms in
the expansion when $n=0$.

Therefore, one obtains,
\begin{eqnarray}
n_{Y,Q,q_{-},0}^{(2)\rho ,\rho _{a},\rho _{b}} &=&\dint dhdh_{a}dh_{b}\frac{1%
}{\rho _{a}^{2}\rho _{b}^{2}}\int_{0}^{Y}\frac{\alpha N_{c}}{\pi }dy\exp %
\left[ \omega \left( n,\nu \right) \left( Y-y\right) +\left( \omega \left(
a\right) +\omega \left( b\right) \right) y\right]  \notag \\
&&\times \frac{b_{n,\nu }b_{n_{a},\nu _{a}}b_{n_{b},\nu _{b}}\left\vert \rho
\rho _{a}\rho _{b}\right\vert }{\left( 2\pi ^{2}\right) ^{3}}E_{Q}^{n,\nu
}\left( \rho \right) E_{q_{a}}^{n_{a},\nu _{a}}\left( \rho _{a}\right)
E_{q_{b}}^{n_{b},\nu _{b}}\left( \rho _{b}\right) g_{3P}\left(
1-h,1-h_{a},1-h_{b}\right)  \notag \\
&&\times 4\pi ^{2}\left( \frac{2}{q_{-}}\right) ^{i+j+k+\overline{i}+%
\overline{j}+\overline{k}-2}\frac{\Gamma \left( k\right) \Gamma \left(
j\right) }{\Gamma \left( k+j\right) }\frac{\Gamma \left( 1-\overline{j}-%
\overline{k}\right) }{\Gamma \left( 1-\overline{j}\right) \Gamma \left( 1-%
\overline{k}\right) }\frac{\Gamma \left( i+j+k-1\right) }{\Gamma \left( 2-%
\overline{i}-\overline{j}-\overline{k}\right) }  \label{n20}
\end{eqnarray}%
where the last subscript of $n_{Y,Q,q_{-},0}^{(2)\rho ,\rho _{a},\rho _{b}}$
means that this result comes from the first term of the expansion. Assuming
that the diffusion approximation is valid which requires $\ln ^{2}\left(
q\rho \right) \ll \frac{14\alpha C_{F}\zeta \left( 3\right) }{\pi }Y$, let
us first evaluate the rapidity-integral. The integral yields
\begin{equation}
\frac{\frac{\alpha N_{c}}{\pi }}{\omega \left( n,\nu \right) -\omega \left(
a\right) -\omega \left( b\right) }\left\{ \exp \left[ \left( \omega \left(
a\right) +\omega \left( b\right) \right) Y\right] -\exp \left[ \omega \left(
n,\nu \right) Y\right] \right\} .
\end{equation}%
For the first term, the saddle point approximation fixes $h_{a}=h_{b}=\frac{1%
}{2}$, $n_{a}=n_{b}=0$. As a result, $n_{Y}^{(2)}$ is then proportional to $%
\exp \left[ 2\left( \alpha _{P}-1\right) Y\right] $. In comparison, for the
second term, $n_{Y}^{(2)}$ is proportional to $\exp \left[ \left( \alpha
_{P}-1\right) Y\right] $ because saddle point approximation fixes $h=\frac{1%
}{2}$ and $n=0$. Therefore, we can drop the second term in later discussions
since it is the next-leading order contribution which is exponentially
suppressed.

In the following, we break the discussions into two parts. The first part is
on the $n=0$ case, and the second part discusses the result when $n\neq 0$.
Usually, we can neglect the contribution from sub-dominant trajectories of
pomerons when rapidity is large enough as we explained in the calculation of
the single pomeron exchange amplitude. Nevertheless, the rapidity interval
of the upper pomeron after integration is infinitesimal in this situation.
Therefore, we have to seriously consider the contribution from sub-dominant
trajectories.

\subsection{$n=0$ part}

$n=0$ part of the amplitude is the only angular-independent part. From
Appendix \ref{saddle}, one uses saddle point approximation to evaluate $%
\dint dh_{a}dh_{b}$ integrals,
\begin{eqnarray}
&&\dint dh_{a}\frac{b_{n_{a},\nu _{a}}\left\vert \rho _{a}\right\vert }{2\pi
^{2}}E_{q_{a}}^{n_{a},\nu _{a}}\left( \rho _{a}\right) \left( \frac{4}{q^{2}}%
\right) ^{\gamma _{a}}\exp \left[ \omega \left( a\right) Y\right]  \notag \\
&\simeq &\frac{1}{2\pi }\frac{\rho _{a}}{q}e\left( \rho _{a},q_{a}\right)
\frac{\exp \left[ \left( \alpha _{P}-1\right) Y\right] \exp \left[ -\frac{%
\pi \ln ^{2}\left( \rho _{a}q_{a}\right) }{28\alpha C_{F}\zeta \left(
3\right) Y}\right] }{\left( 7\alpha \zeta \left( 3\right) C_{F}Y\right)
^{3/2}},
\end{eqnarray}%
where $\gamma _{a}=\frac{1}{2}\left( h_{a}+\overline{h}_{a}\right) =\frac{1}{%
2}+i\nu _{a}$. In the end, there is only $\dint dh$ left in the expression,
\begin{eqnarray}
n_{Y,Q,q_{-},0}^{(2)\rho ,\rho _{a},\rho _{b}} &\simeq &\dint dh\frac{1}{%
2\rho _{a}^{2}\rho _{b}^{2}}\frac{1}{2\chi \left( \frac{1}{2}\right) -\chi
\left( \gamma \right) }\frac{b_{0,\nu }\left\vert \rho \right\vert }{2\pi
^{2}}E_{Q}^{0,\nu }\left( \rho \right) g_{3P}\left( 1-\gamma ,\frac{1}{2},%
\frac{1}{2}\right)  \notag \\
&&\times 4\pi ^{2}\left( \frac{4}{q_{-}^{2}}\right) ^{\gamma -1}\frac{\Gamma
^{3}\left( \gamma \right) \Gamma \left( 1-2\gamma \right) }{\Gamma \left(
2\gamma \right) \Gamma ^{3}\left( 1-\gamma \right) }  \notag \\
&&\times \frac{1}{4\pi ^{2}}\frac{\rho _{a}\rho _{b}}{q_{-}^{2}}e\left( \rho
_{a},q_{a}\right) e\left( \rho _{b},q_{b}\right) \frac{\exp \left[ 2\left(
\alpha _{P}-1\right) Y\right] \exp \left[ -\frac{\pi \left( \ln ^{2}\left(
q_{a}\rho _{a}\right) +\ln ^{2}\left( q_{b}\rho _{b}\right) \right) }{%
28\alpha C_{F}\zeta \left( 3\right) Y}\right] }{\left( 7\alpha \zeta \left(
3\right) C_{F}Y\right) ^{3}}
\end{eqnarray}%
Let us focus on the $dh$-integral (first two lines of the above expression):%
\begin{eqnarray}
&&\frac{4\pi ^{2}}{2\rho _{a}^{2}\rho _{b}^{2}}\dint dh\frac{1}{2\chi \left(
\frac{1}{2}\right) -\chi \left( \gamma \right) }\frac{b_{0,\nu }\left\vert
\rho \right\vert }{2\pi ^{2}}E_{Q}^{0,\nu }\left( \rho \right) g_{3P}\left(
1-\gamma ,\frac{1}{2},\frac{1}{2}\right) \left( \frac{4}{q_{-}^{2}}\right)
^{\gamma -1}\frac{\Gamma ^{3}\left( \gamma \right) \Gamma \left( 1-2\gamma
\right) }{\Gamma \left( 2\gamma \right) \Gamma ^{3}\left( 1-\gamma \right) },
\notag \\
&=&\frac{2}{\rho _{a}^{2}\rho _{b}^{2}}\int \frac{d\gamma }{2\pi i}\frac{%
\left\vert \rho \right\vert }{2\chi \left( \frac{1}{2}\right) -\chi \left(
\gamma \right) }E_{Q}^{n,\nu }\left( \rho \right) g_{3P}\left( 1-\gamma ,%
\frac{1}{2},\frac{1}{2}\right) \left( \frac{4}{q_{-}^{2}}\right) ^{\gamma -1}%
\frac{\Gamma \left( \gamma \right) }{\Gamma \left( 1-\gamma \right) }.
\end{eqnarray}
In $\rho Q\ll 1$ limit\cite{Navelet:1997xn},
\begin{equation}
E_{Q}^{0,\nu }\left( \rho \right) \simeq \rho ^{1-2\gamma }-\frac{Q^{2\gamma
-1}\Gamma ^{2}\left( \frac{3}{2}-\gamma \right) 2^{6-12\gamma }}{\Gamma
^{2}\left( \frac{1}{2}+\gamma \right) }\left( \rho Q\right) ^{2\gamma -1}.
\end{equation}
Therefore, the integral can be cast into the form
\begin{eqnarray}
dh-integral &\simeq &\frac{2}{\rho _{a}^{2}\rho _{b}^{2}}\int_{\frac{1}{2}%
-i\infty }^{\frac{1}{2}+i\infty }\frac{d\gamma }{2\pi i}\frac{1}{2\chi
\left( \frac{1}{2}\right) -\chi \left( \gamma \right) }g_{3P}\left( 1-\gamma
,\frac{1}{2},\frac{1}{2}\right) \frac{\Gamma \left( \gamma \right) }{\Gamma
\left( 1-\gamma \right) }\left( \frac{4}{\rho ^{2}q_{-}^{2}}\right) ^{\gamma
-1}  \notag \\
&&-\frac{2}{\rho _{a}^{2}\rho _{b}^{2}}\int_{\frac{1}{2}-i\infty }^{\frac{1}{%
2}+i\infty }\frac{d\gamma }{2\pi i}\frac{1}{2\chi \left( \frac{1}{2}\right)
-\chi \left( \gamma \right) }g_{3P}\left( 1-\gamma ,\frac{1}{2},\frac{1}{2}%
\right) \frac{\Gamma \left( \gamma \right) }{\Gamma \left( 1-\gamma \right) }
\notag \\
&&\times \frac{\Gamma ^{2}\left( \frac{3}{2}-\gamma \right) 2^{6-12\gamma }}{%
\Gamma ^{2}\left( \frac{1}{2}+\gamma \right) }\left( \frac{4Q^{2}}{q_{-}^{2}}%
\right) ^{2\gamma -1}\left( \frac{\rho ^{2}q_{-}^{2}}{4}\right) ^{\gamma }
\end{eqnarray}

\begin{figure}[tbp]
\begin{center}
\includegraphics[width=5cm]{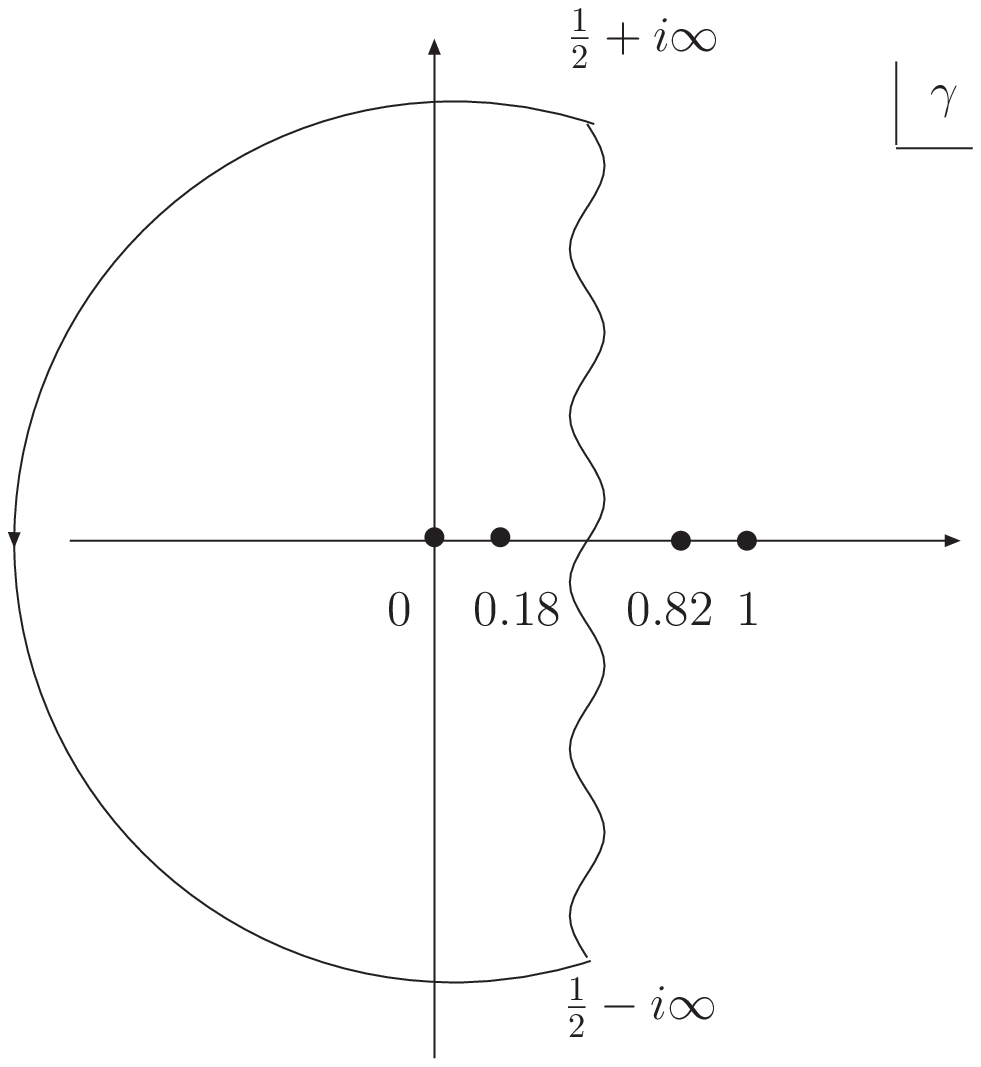} \includegraphics[width=5cm]{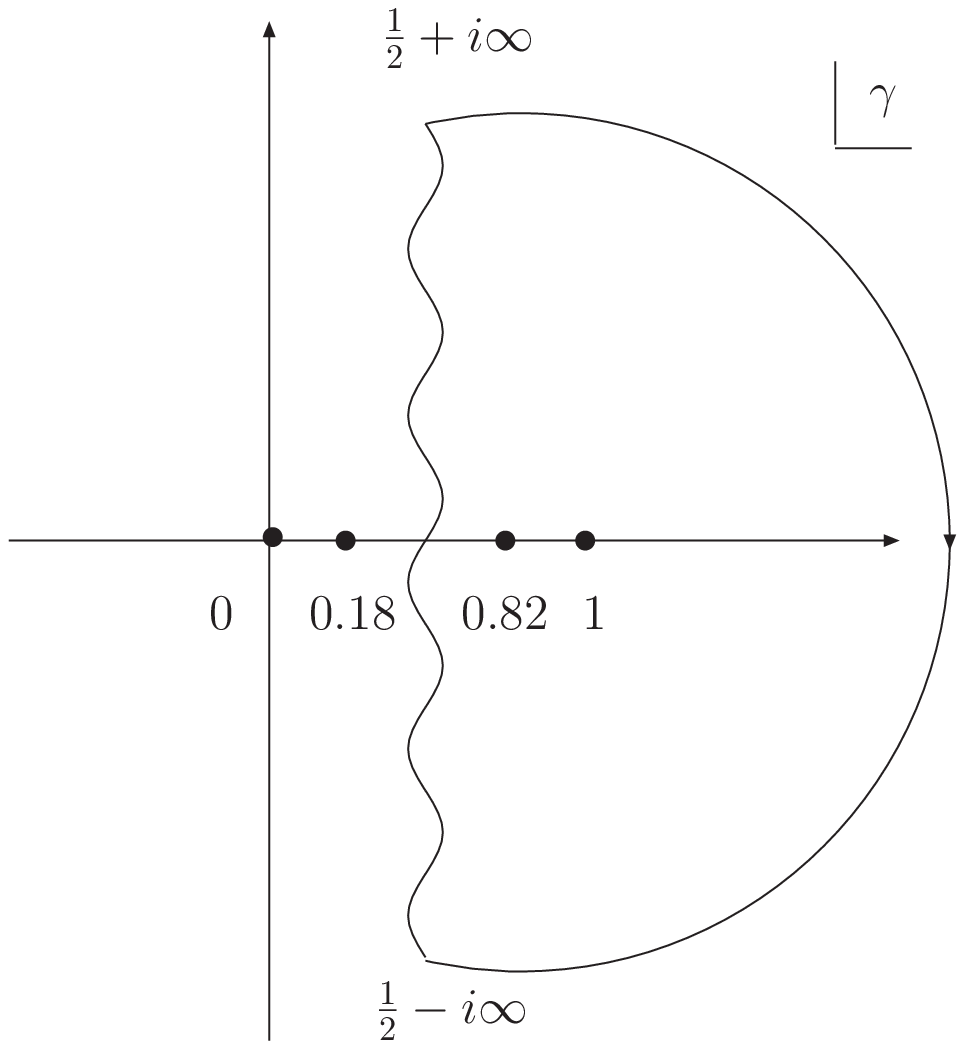}
\end{center}
\caption[*]{Illustration of the contours in complex $\protect\gamma $ plane
for the computation of $n_{Y}^{(2)}$ when $n=0$. }
\label{con}
\end{figure}
According to the residue theorem, the contour integral equals to the sum of
all residues of poles enclosed by the contour. In the $\rho ^{2}q_{-}^{2}\ll
1$ limit, the first poles to the left and right of the vertical contour are
the dominant contribution. The positions of these two poles are determined
by the equation $2\chi \left( \frac{1}{2}\right) -\chi \left( \gamma \right)
=0$, and they are $\gamma _{0}^{\left( 2\right) \ast }=0.18$ and $1-\gamma
_{0}^{\left( 2\right) \ast }=0.82$. For the first integral(see the left
graph of Fig. \ref{con}), we should close the contour to the left to $%
-\infty $ and compute the residue of the pole at $\gamma =0.18$; while for
the second integral(see the right graph of Fig. \ref{con}), we should close
the contour to the right to $\infty $ and compute the residue at $\gamma
=0.82$. Eventually, up to leading order precision, one reaches the result
\begin{eqnarray}
n_{Y,Q,q_{a},q_{b},0}^{(2)\rho ,\rho _{a},\rho _{b}} &=&\pi ^{2}\delta
^{\left( 2\right) }\left( Q+q_{a}+q_{b}\right) n_{Y,Q,q_{-},0}^{(2)\rho
,\rho _{a},\rho _{b}}  \notag \\
&\simeq &\frac{1}{2\rho _{a}q_{-}\rho _{b}q_{-}}\left( \frac{\rho
^{2}q_{-}^{2}}{4}\right) ^{1-\gamma _{0}^{\left( 2\right) \ast }}\delta
^{\left( 2\right) }\left( Q+q_{a}+q_{b}\right) \frac{g_{3P}\left( 1-\gamma
_{0}^{\left( 2\right) \ast },\frac{1}{2},\frac{1}{2}\right) }{\left\vert
\chi ^{\prime }\left( \gamma _{0}^{\left( 2\right) \ast }\right) \right\vert
}  \notag \\
&&\times \left[ \frac{\Gamma \left( \gamma _{0}^{\left( 2\right) \ast
}\right) }{\Gamma \left( 1-\gamma _{0}^{\left( 2\right) \ast }\right) }-%
\frac{\Gamma \left( 1-\gamma _{0}^{\left( 2\right) \ast }\right) }{\Gamma
\left( \gamma _{0}^{\left( 2\right) \ast }\right) }\frac{\Gamma ^{2}\left(
\frac{1}{2}+\gamma _{0}^{\left( 2\right) \ast }\right) 2^{-6+12\gamma
_{0}^{\left( 2\right) \ast }}}{\Gamma ^{2}\left( \frac{3}{2}-\gamma
_{0}^{\left( 2\right) \ast }\right) }\left( \frac{4Q^{2}}{q_{-}^{2}}\right)
^{-2\gamma _{0}^{\left( 2\right) \ast }+1}\right]  \notag \\
&&\times e\left( \rho _{a},q_{a}\right) e\left( \rho _{b},q_{b}\right) \frac{%
\exp \left[ 2\left( \alpha _{P}-1\right) Y\right] \exp \left[ -\frac{\pi
\left( \ln ^{2}\left( q_{a}\rho _{a}\right) +\ln ^{2}\left( q_{b}\rho
_{b}\right) \right) }{28\alpha C_{F}\zeta \left( 3\right) Y}\right] }{\left(
7\alpha \zeta \left( 3\right) C_{F}Y\right) ^{3}}.  \label{n2}
\end{eqnarray}
In this leading order calculation of $1\Rightarrow 2$ amplitude, we push the
rapidity integration to its upper limit $Y$ which leaves infinitesimal
rapidity for the upper pomeron. As a result, the anomalous dimensions is
determined by the dynamical pole introduced by the rapidity integration.
This result agrees with eq.(44) in ref.\cite{Mueller:1994jq}.
\begin{eqnarray}
n_{2}\left( Y,x_{01},q\right) &=&\frac{\pi x_{01}^{2}\exp \left[ 2\left(
\alpha _{P}-1\right) Y\right] }{8\left( 7\alpha \zeta \left( 3\right)
C_{F}Y\right) ^{3}}\int_{-\infty}^{+\infty} \frac{d\nu }{2\pi}\left(
qx_{01}\right) ^{2i\nu -1}\frac{V_{\nu }}{4\ln 2-\chi \left( \nu \right) } \\
&\simeq &\frac{\pi \exp \left[ 2\left( \alpha _{P}-1\right) Y\right] }{
8\left( 7\alpha \zeta \left( 3\right) C_{F}Y\right) ^{3}}\frac{\left(
q^{2}x_{01}^{2}\right) ^{0.82}}{q^{2}}\frac{V_{\nu _{0}^{(2)\ast }}}{
\left\vert \chi ^{\prime }\left( \nu _{0}^{(2)\ast }\right) \right\vert },
\end{eqnarray}%
where $V_{\nu }$ is defined by
\begin{eqnarray}
\int_{-\infty}^{+\infty} \frac{d\nu }{2\pi }\left( qx_{01}\right) ^{2i\nu
-1}V_{\nu }=\int \frac{ d^{2}x_{2}}{x_{01}x_{12}}J_{0}\left( \frac{1}{2}%
qx_{01}\right) e\left( x_{12},q\right) e\left( x_{02},q\right) ,
\label{mellin}
\end{eqnarray}
and
\begin{equation}
\gamma _{0}^{\left( 2\right) \ast \prime }=\frac{1}{2}+i\nu_{0} ^{\left(
2\right) \ast }=0.82.
\end{equation}%
In reaching the final result, we have assumed that $V_{\nu }$ has no pole
before $\gamma _{0}^{\left( 2\right) \ast \prime}$. The proof of this
assumption is explicitly provided in appendix \ref{v}.

\subsection{$n\neq 0$ part}

\begin{figure}[tbp]
\begin{center}
\includegraphics[width=5cm]{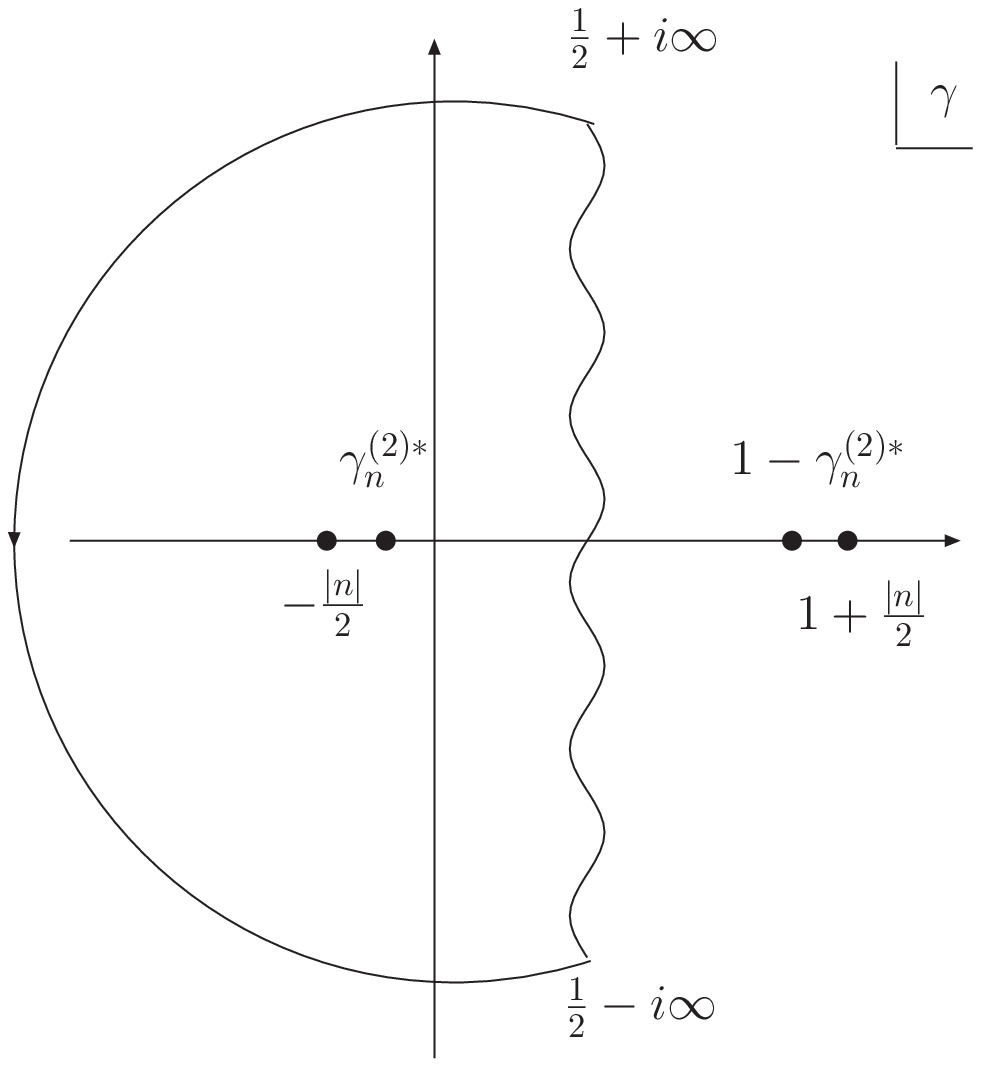} \includegraphics[width=5cm]{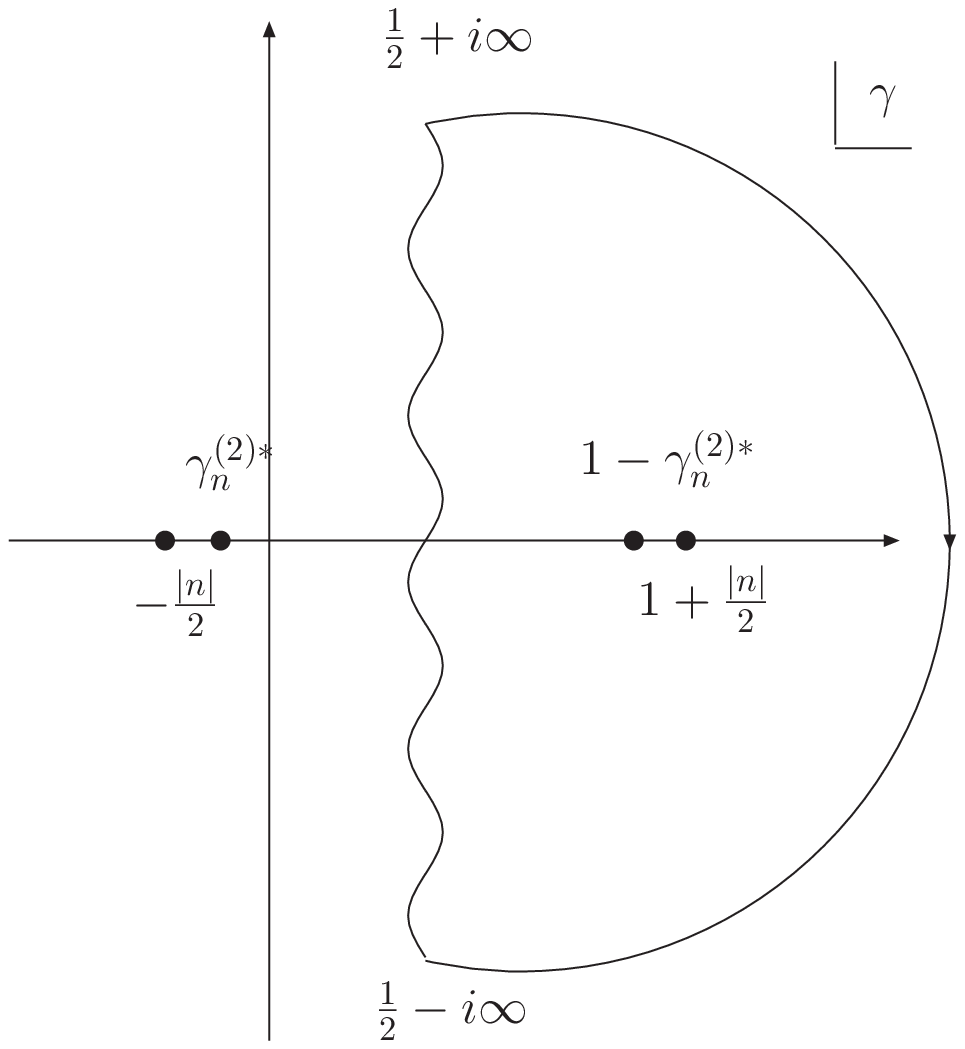}
\end{center}
\caption[*]{Illustration of the contours in complex $\protect\gamma $ plane
for the computation of $n_{Y}^{(2)}$ when $n \neq 0$. }
\label{con34}
\end{figure}

$n\neq 0$ part of the amplitude is angular-dependent, and it vanishes after
averaging over the angle of initial dipole orientation. This part of
amplitude comes from the subdominant trajectories of the upper pomeron. As
discussed in Appendix \ref{general}, we should restrict ourselves to the
forward scattering case in which $Q=0$, since we are unable to prove that
the whole discussion can be generalized to arbitrary momenta. According to
the discussion above, one can easily obtain the expression for $%
n_{Y}^{\left( 2\right) }$ for nonzero value of $n$ in forward scattering.
\begin{eqnarray}
\left. n_{Y,Q=0,q_{-}}^{(2)\rho ,\rho _{a},\rho _{b}}\right\vert
_{\left\vert n\right\vert >0} &\simeq &\dint_{n\neq 0}dh\frac{b_{n,\nu
}\left\vert \rho \right\vert }{\rho _{a}^{2}\rho _{b}^{2}}\frac{1}{2\chi
\left( \frac{1}{2}\right) -\chi \left( n,\gamma \right) }E_{0}^{n,\nu
}\left( \rho \right) g_{3P}\left( 1-h,\frac{1}{2},\frac{1}{2}\right) \left(
\frac{4}{q_{-}^{2}}\right) ^{\gamma -1}\frac{\Gamma ^{3}\left( h\right)
\Gamma \left( 1-2\overline{h}\right) }{\Gamma \left( 2h\right) \Gamma
^{3}\left( 1-\overline{h}\right) }  \notag \\
&&\times \frac{1}{4\pi ^{2}}\frac{\rho _{a}\rho _{b}}{q_{-}^{2}}e\left( \rho
_{a},q_{a}\right) e\left( \rho _{b},q_{b}\right) \frac{\exp \left[ 2\left(
\alpha _{P}-1\right) Y\right] \exp \left[ -\frac{\pi \left( \ln ^{2}\left(
q_{a}\rho _{a}\right) +\ln ^{2}\left( q_{b}\rho _{b}\right) \right) }{%
28\alpha C_{F}\zeta \left( 3\right) Y}\right] }{\left( 7\alpha \zeta \left(
3\right) C_{F}Y\right) ^{3}}
\end{eqnarray}%
Changing $\dint_{n\neq 0}dh$ into $\int \frac{d\gamma }{2\pi i}$, and using
the identity $g_{3P}\left( 1-h,\frac{1}{2},\frac{1}{2}\right) =g_{3P}\left(
1-\overline{h},\frac{1}{2},\frac{1}{2}\right) $ which we proved in Appendix %
\ref{g3p}, along with the detailed derivation in Appendix \ref{highn}, one
can reach
\begin{eqnarray}
\left. n_{Y,Q=0,q_{-}}^{(2)\rho ,\rho _{a},\rho _{b}}\right\vert
_{\left\vert n\right\vert >0} &\simeq &\sum_{n=1}^{\infty }\frac{2\left\vert
\rho \right\vert }{\rho _{a}^{2}\rho _{b}^{2}}\int_{\frac{1}{2}-i\infty }^{%
\frac{1}{2}+i\infty }\frac{d\gamma }{2\pi i}\frac{1}{2\chi \left( \frac{1}{2}%
\right) -\chi \left( n,\gamma \right) }\frac{\Gamma \left( \gamma +\frac{%
\left\vert n\right\vert }{2}\right) }{\Gamma \left( 1+\frac{\left\vert
n\right\vert }{2}-\gamma \right) }\left( \frac{4}{q_{-}^{2}}\right) ^{\gamma
-1}  \notag \\
&&\times i^{\left\vert n\right\vert }\left[ E_{0}^{n,\nu }\left( \rho
\right) +E_{0}^{-n,\nu }\left( \rho \right) \right] g_{3P}\left( 1-h,\frac{1%
}{2},\frac{1}{2}\right)  \notag \\
&&\times \frac{1}{4\pi ^{2}}\frac{\rho _{a}\rho _{b}}{q_{-}^{2}}e\left( \rho
_{a},q_{a}\right) e\left( \rho _{b},q_{b}\right) \frac{\exp \left[ 2\left(
\alpha _{P}-1\right) Y\right] \exp \left[ -\frac{\pi \left( \ln ^{2}\left(
q_{a}\rho _{a}\right) +\ln ^{2}\left( q_{b}\rho _{b}\right) \right) }{%
28\alpha C_{F}\zeta \left( 3\right) Y}\right] }{\left( 7\alpha \zeta \left(
3\right) C_{F}Y\right) ^{3}}
\end{eqnarray}%
In $\rho Q\ll 1$ limit\cite{Navelet:1997xn},
\begin{eqnarray}
E_{Q}^{n,\nu }\left( \rho \right) &=&\overline{Q}^{i\nu -n/2}Q^{i\nu
+n/2}2^{-6i\nu }\Gamma \left( 1-i\nu +\frac{\left\vert n\right\vert }{2}%
\right) \Gamma \left( 1-i\nu -\frac{\left\vert n\right\vert }{2}\right) \\
&&\times \left[ J_{\frac{n}{2}-i\nu }\left( \frac{\overline{Q}\rho }{4}%
\right) J_{-\frac{n}{2}-i\nu }\left( \frac{\overline{\rho }Q}{4}\right)
-\left( -1\right) ^{n}J_{-\frac{n}{2}+i\nu }\left( \frac{\overline{Q}\rho }{4%
}\right) J_{\frac{n}{2}+i\nu }\left( \frac{\overline{\rho }Q}{4}\right) %
\right]  \notag \\
&\simeq &\left( \frac{\rho }{\overline{\rho }}\right) ^{\frac{n}{2}}\rho
^{1-2\gamma }-\left( -1\right) ^{n}\frac{Q^{2\gamma -1}\Gamma \left( \frac{3%
}{2}-\gamma +\frac{n}{2}\right) \Gamma \left( \frac{3}{2}-\gamma -\frac{n}{2}%
\right) 2^{6-12\gamma }}{\Gamma \left( \frac{1}{2}+\gamma +\frac{n}{2}%
\right) \Gamma \left( \frac{1}{2}+\gamma -\frac{n}{2}\right) }\left( \frac{Q%
}{\overline{Q}}\right) ^{n}\left( \frac{\rho }{\overline{\rho }}\right) ^{-%
\frac{n}{2}}\left( \rho Q\right) ^{2\gamma -1}.
\end{eqnarray}%
where $\left( \frac{\rho }{\overline{\rho }}\right) ^{\frac{n}{2}}$ and $%
\left( \frac{\rho }{\overline{\rho }}\right) ^{-\frac{n}{2}}$ parts give the
angular dependence, and vanish after averaging. Thus, $E_{0}^{n,\nu }\left(
\rho \right) +E_{0}^{-n,\nu }\left( \rho \right) =\left[ \left( \frac{\rho }{%
\overline{\rho }}\right) ^{\frac{n}{2}}+\left( \frac{\rho }{\overline{\rho }}%
\right) ^{-\frac{n}{2}}\right] \rho ^{1-2\gamma }$. Similarly, one has to
examine the pole structure of the integrand of the $\gamma $ integral.
Assuming that $g_{3P}\left( 1-h,\frac{1}{2},\frac{1}{2}\right) $ does not
contribute any new singularity in the domain $\left( -\frac{\left\vert
n\right\vert }{2},\frac{\left\vert n\right\vert }{2}+1\right) $(see Appendix %
\ref{g3p}), noting that $\chi \left( \gamma ,n\right) $ is analytic in the
strip $-\frac{\left\vert n\right\vert }{2}<\func{Re}\nu <\frac{\left\vert
n\right\vert }{2}+1$, and defining $\gamma _{n}^{\left( 2\right) \ast }$ as
the solution to the equations $2\chi \left( \frac{1}{2}\right) -\chi \left(
n,\gamma \right) =0$ in the domain $\left( -\frac{\left\vert n\right\vert }{2%
},\frac{1}{2}\right) $(see the left graph of Fig.\ref{con34}), one reaches
\begin{equation}
\left. n_{Y,Q=0,q_{-}}^{(2)\rho ,\rho _{a},\rho _{b}}\right\vert _{n}\propto
\frac{\left( \rho ^{2}q_{-}^{2}\right) ^{1-\gamma _{n}^{\left( 2\right) \ast
}}}{\rho _{a}q_{-}\rho _{b}q_{-}}e\left( \rho _{a},q_{a}\right) e\left( \rho
_{b},q_{b}\right) \frac{\exp \left[ 2\left( \alpha _{P}-1\right) Y\right]
\exp \left[ -\frac{\pi \left( \ln ^{2}\left( q_{a}\rho _{a}\right) +\ln
^{2}\left( q_{b}\rho _{b}\right) \right) }{28\alpha C_{F}\zeta \left(
3\right) Y}\right] }{\left( 7\alpha \zeta \left( 3\right) C_{F}Y\right) ^{3}}
\end{equation}%
by closing the contour to the left to $-\infty $ and collecting the residue
at $\gamma =\gamma _{n}^{\left( 2\right) \ast }$ (other residues are
suppressed by factors of $\rho ^{2}q_{-}^{2}$). It seems that the dynamical
pole introduced by the rapidity integrations always comes before other
poles. Here we conjecture that, although we can not prove, $\left.
n_{Y,Q=0,q_{-}}^{(2)\rho ,\rho _{a},\rho _{b}}\right\vert _{n}\propto \frac{%
\left( \rho ^{2}q_{-}^{2}\right) ^{1-\gamma _{n}^{\left( 2\right) \ast }}}{%
\rho _{a}q_{-}\rho _{b}q_{-}}$ still holds even for nonzero momenta of $Q $
by picking up poles at $\gamma _{n}^{\left( 2\right) \ast }$ or $1-\gamma
_{n}^{\left( 2\right) \ast }$ depending on how the contour is closed(as
shown in both graphs of Fig.~\ref{con34}).

Comparing this result to the result in the case of $n=0$, one can spot that $%
n_{Y}^{(2)}$ with nonzero $n$ is suppressed by factors of $\rho
^{2}q_{-}^{2} $ in the $\rho ^{2}q_{-}^{2}\ll 1$ limit since $1-\gamma
_{n}^{\left( 2\right) \ast }\simeq 1+\frac{\left\vert n\right\vert }{2}%
-\epsilon _{n}$, where $\epsilon _{n}$ is a small positive number.
Therefore, the sub-dominant trajectories can be neglected in the case of
forward scattering computation. It is our conjecture that they can also be
neglected in the non-forward scattering case. This result is new.

To summarize, we have shown that the dipole pair density in momentum space
scales as $\left( \rho ^{2}q^{2}\right) ^{1-\gamma _{0}^{\left( 2\right)
\ast }}$ with respect to the onium size in $\rho ^{2}q^{2}\ll 1$ limit. This
indicates that the anomalous dimension of dipole pair density is equal to $%
0.18$. Similar result can also be found in ref.~(\cite{Navelet:2002zz}).
Qualitatively, this result is easy to understand according to the discussion
in Section~\ref{single}. The anomalous dimension is determined by the upper
pomeron which connects the triple pomeron vertex to the initial onium as
seen in fig.~\ref{n2graph}. The rapidity length of the upper pomeron is
infinitesimal after rapidity integration. The anomalous dimension should
approach zero in small rapidity limit or DGLAP limit ($\rho ^{2}q^{2}\ll 1$
implies the collinear ordering and the evolution in virtuality). In the case
of onium-onium scattering, the double pomeron exchange amplitude between two
onia with sizes $\rho $ and $\rho ^{\prime }$ then scales as
\begin{equation}
F^{\left( 2\right) }\left( \rho ,\rho ^{\prime },q,Y\right) \propto \left(
q^{2}\rho ^{2}\right) ^{0.82}\left( q^{2}\rho ^{\prime 2}\right)
^{0.82}\left\{ \alpha ^{2}\frac{\exp \left[ \left( \alpha _{P}-1\right) Y%
\right] }{\left( 7\alpha \zeta \left( 3\right) C_{F}Y\right) ^{3/2}}\right\}
^{2}.
\end{equation}

\section{The $1\Rightarrow 3$ amplitude and Triple pomeron exchange.}

\label{triple3}

\begin{figure}[tbp]
\begin{center}
\includegraphics[width=6cm]{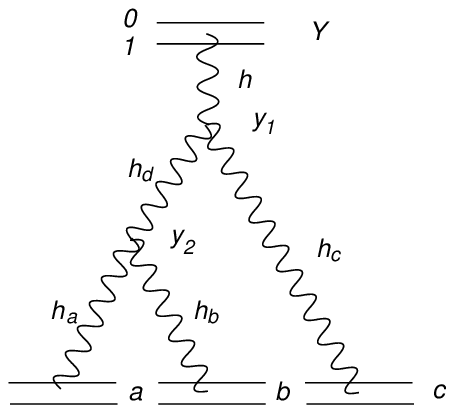} \includegraphics[width=6.5cm]{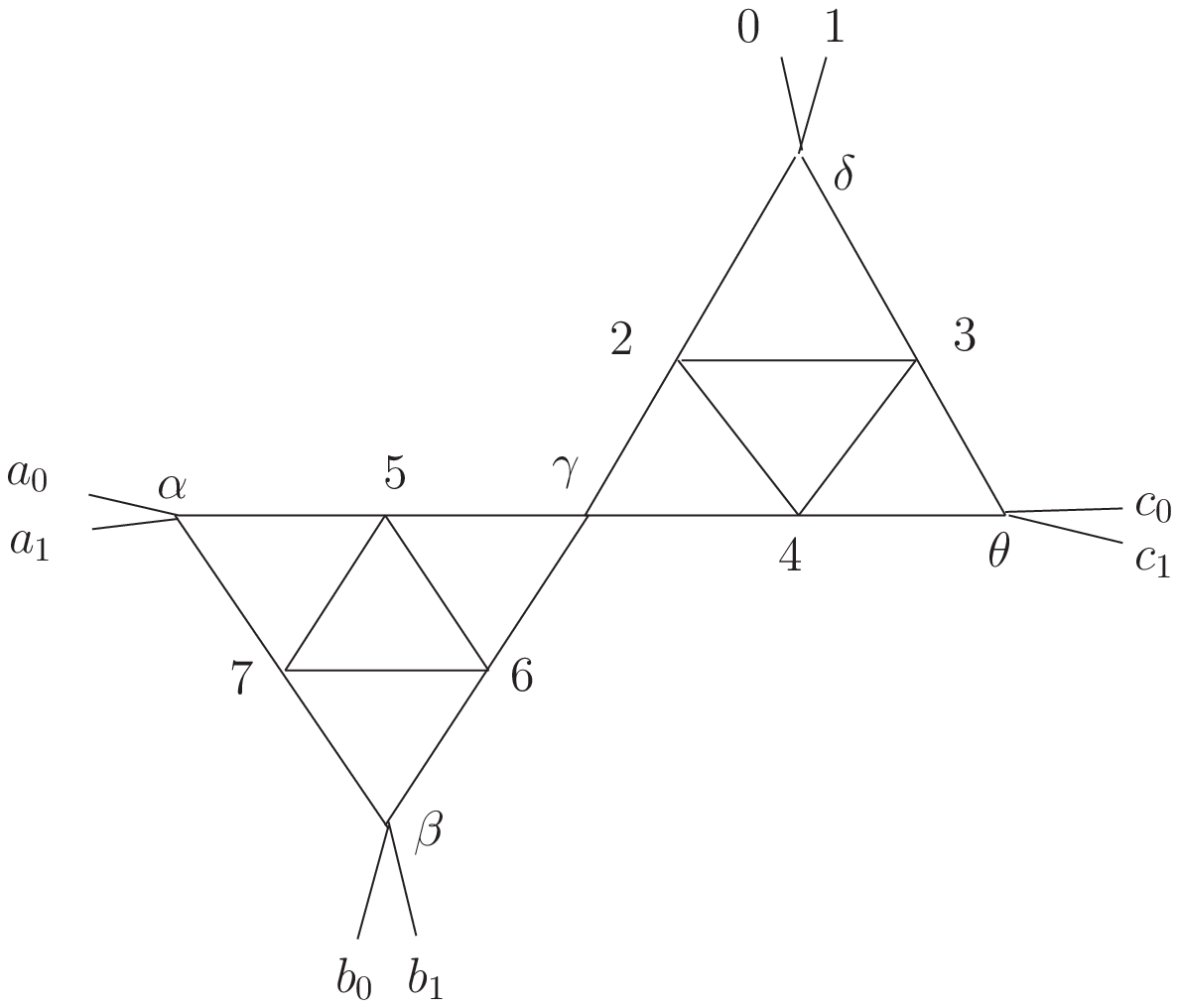}
\end{center}
\caption[*]{Dipole triplet density $n_{Y}^{3}$ in an onium state with
initial size $\protect\rho =\protect\rho _{0}-\protect\rho _{1}$ and three
children dipoles having transverse separation $\protect\rho _{a}$, $\protect%
\rho _{b}$ and $\protect\rho _{c}$. The left graph shows the rapidity
structure of $n_{3}$, and the right graphs indicates the graphical
representation of two successive triple pomeron vertices.}
\label{n31}
\end{figure}
As shown in Fig.~\ref{n31}, the $1\Rightarrow 3$ amplitude is defined as the
dipole triplet density in an onium state. It is easy to generalize the $%
1\Rightarrow 2$ amplitude to the $1\Rightarrow 3$ amplitude by adding one
more triple-pomeron vertex. This amplitude corresponds to the graph in which
one ancestor pomeron splits into two pomersons and one of the two descendent
pomerons again splits into another two pomerons. As discussed in ref.~\cite%
{Braun:1997nu}, the amplitude of this graph is different from the $%
1\Rightarrow 3$ amplitude introduced in ref.~\cite{Peschanski:1997yx} which
involves a non-local $1\Rightarrow 3$ pomeron vertex(see also ref.~\cite%
{Janik:1999fk}). As seen in ref.~\cite{Peschanski:1997yx}, an initial
pomeron can split into $p$ ($p>1$) pomerons simultaneously via a non-local $%
1\Rightarrow p$ pomeron vertex. The use of the non-local $1\Rightarrow p$
pomeron vertex ($p>1$) is still under debate\cite{Braun:1997nu}. It seems
more natural that the theory only requires one type of vertex (triple
pomeron vertex) instead of infinite number of different vertices. Therefore,
the $1\Rightarrow 3$ amplitude reads,
\begin{eqnarray}
&&n_{Y}^{(3)}\left( \rho _{0}\rho _{1};\rho _{a_{0}}\rho _{a_{1}},\rho
_{b_{0}}\rho _{b_{1}},\rho _{c_{0}}\rho _{c_{1}}\right)  \notag \\
&=&\frac{1}{\rho _{a}^{2}\rho _{b}^{2}\rho _{c}^{2}}\dint
dhdh_{a}dh_{b}dh_{c}dh_{d}\dint d^{2}\rho _{\alpha }d^{2}\rho _{\beta
}d^{2}\rho _{\gamma }d^{2}\rho _{\theta }d^{2}\rho _{\delta }  \notag \\
&&\times \dint \frac{d\omega }{2\pi i}\frac{e^{\omega Y}}{\omega -\omega
\left( n,\nu \right) }\frac{\frac{\alpha N_{c}}{\pi }}{\omega -\omega \left(
c\right) -\omega \left( d\right) }\frac{\frac{\alpha N_{c}}{\pi }}{\omega
-\omega \left( a\right) -\omega \left( b\right) -\omega \left( c\right) }
\notag \\
&&\times E^{h,\overline{h}}\left( \rho _{0\delta }\rho _{1\delta }\right)
E^{h_{c},\overline{h}_{c}}\left( \rho _{c_{0}\theta }\rho _{c_{1}\theta
}\right) E^{h_{a},\overline{h}_{a}}\left( \rho _{a_{0}\alpha }\rho
_{a_{1}\alpha }\right) E^{h_{b},\overline{h}_{b}}\left( \rho _{b_{0}\beta
}\rho _{b_{1}\beta }\right)  \notag \\
&&\times \dint \frac{d^{2}\rho _{2}d^{2}\rho _{3}d^{2}\rho _{4}}{\left\vert
\rho _{23}\rho _{34}\rho _{42}\right\vert ^{2}}E^{h,\overline{h}\ast }\left(
\rho _{2\delta }\rho _{3\delta }\right) E^{h_{d},\overline{h}_{d}}\left(
\rho _{2\gamma }\rho _{4\gamma }\right) \overline{E}^{h_{c},\overline{h}%
_{c}\ast }\left( \rho _{3\theta }\rho _{4\theta }\right)  \notag \\
&&\times \dint \frac{d^{2}\rho _{5}d^{2}\rho _{6}d^{2}\rho _{7}}{\left\vert
\rho _{56}\rho _{57}\rho _{67}\right\vert ^{2}}E^{h_{d},\overline{h}_{d}\ast
}\left( \rho _{5\gamma }\rho _{6\gamma }\right) E^{h_{a},\overline{h}%
_{a}\ast }\left( \rho _{5\alpha }\rho _{7\alpha }\right) \overline{E}^{h_{b},%
\overline{h}_{b}\ast }\left( \rho _{6\beta }\rho _{7\beta }\right)
\end{eqnarray}%
Following the same procedure as in the last section, first of all, we define
\begin{eqnarray}
\overline{R}_{1,\delta ,\gamma ,\theta }^{h,1-h_{a},h_{b}} &=&R_{1,\delta
,\gamma ,\theta }^{1-h,h_{d},1-h_{c}}  \notag \\
&=&\dint \frac{d^{2}\rho _{2}d^{2}\rho _{3}d^{2}\rho _{4}}{\left\vert \rho
_{23}\rho _{34}\rho _{42}\right\vert ^{2}}E^{h,\overline{h}\ast }\left( \rho
_{2\delta }\rho _{3\delta }\right) E^{h_{d},\overline{h}_{d}}\left( \rho
_{2\gamma }\rho _{4\gamma }\right) \overline{E}^{h_{c},\overline{h}_{c}\ast
}\left( \rho _{3\theta }\rho _{4\theta }\right) ,
\end{eqnarray}%
and%
\begin{eqnarray}
\overline{R}_{2,\gamma ,\alpha ,\beta }^{h_{d},h_{a},h_{b}} &=&R_{2,\gamma
,\alpha ,\beta }^{1-h_{d},1-h_{a},1-h_{b}}  \notag \\
&=&\dint \frac{d^{2}\rho _{5}d^{2}\rho _{6}d^{2}\rho _{7}}{\left\vert \rho
_{56}\rho _{57}\rho _{67}\right\vert ^{2}}E^{h_{d},\overline{h}_{d}\ast
}\left( \rho _{5\gamma }\rho _{6\gamma }\right) E^{h_{a},\overline{h}%
_{a}\ast }\left( \rho _{5\alpha }\rho _{7\alpha }\right) \overline{E}^{h_{b},%
\overline{h}_{b}\ast }\left( \rho _{6\beta }\rho _{7\beta }\right) .
\end{eqnarray}%
Using $SL\left( 2,C\right) $ transformation, one can easily reach,
\begin{eqnarray}
R_{1,\delta ,\gamma ,\theta }^{1-h,h_{d},1-h_{c}} &=&g_{3P}\left(
1-h,h_{d},1-h_{c}\right) \left( \rho _{\gamma \theta }^{-1+i^{\prime }}\rho
_{\theta \delta }^{-1+j^{\prime }}\rho _{\delta \gamma }^{-1+k^{\prime
}}\right) \left( \overline{\rho }_{\gamma \theta }^{-1+\overline{i}^{\prime
}}\overline{\rho }_{\theta \delta }^{-1+\overline{j}^{\prime }}\overline{%
\rho }_{\delta \gamma }^{-1+\overline{k}^{\prime }}\right) , \\
R_{2,\gamma ,\alpha ,\beta }^{1-h_{d},1-h_{a},1-h_{b}} &=&g_{3P}\left(
1-h_{d},1-h_{a},1-h_{b}\right) \left( \rho _{\alpha \beta }^{-1+i^{\prime
\prime }}\rho _{\beta \gamma }^{-1+j^{\prime \prime }}\rho _{\gamma \alpha
}^{-1+k^{\prime \prime }}\right) \left( \overline{\rho }_{\alpha \beta }^{-1+%
\overline{i}^{\prime \prime }}\overline{\rho }_{\beta \gamma }^{-1+\overline{%
j}^{\prime \prime }}\overline{\rho }_{\gamma \alpha }^{-1+\overline{k}%
^{\prime \prime }}\right) ,
\end{eqnarray}%
where%
\begin{eqnarray}
i^{\prime } &=&-h+1-h_{d}+h_{c}, \\
j^{\prime } &=&-1+h_{d}+h_{c}+h, \\
k^{\prime } &=&-h_{c}+1-h_{d}+h,
\end{eqnarray}%
and
\begin{eqnarray}
i^{\prime \prime } &=&-h_{d}+h_{a}+h_{b}, \\
j^{\prime \prime } &=&-h_{a}+h_{b}+h_{d}, \\
k^{\prime \prime } &=&-h_{b}+h_{d}+h_{a}.
\end{eqnarray}%
Moreover, in momentum space, we have
\begin{eqnarray}
n_{Y,Q,q_{a},q_{b},q_{c}}^{(3)\rho ,\rho _{a},\rho _{b},\rho _{c}} &=&\int
d^{2}rd^{2}r_{a}d^{2}r_{b}d^{2}r_{c}e^{iQr+q_{a}r_{a}+iq_{b}r_{b}+iq_{c}r_{c}}n_{Y}^{(3)}\left( \rho _{0}\rho _{1};\rho _{a_{0}}\rho _{a_{1}},\rho _{b_{0}}\rho _{b_{1}},\rho _{c_{0}}\rho _{c_{1}}\right)
\notag \\
&=&\dint dhdh_{a}dh_{b}dh_{c}dh_{d}\dint \frac{d\omega }{2\pi i}\frac{%
e^{\omega Y}}{\omega -\omega \left( n,\nu \right) }\frac{\frac{\alpha N_{c}}{%
\pi }}{\omega -\omega \left( c\right) -\omega \left( d\right) }\frac{\frac{%
\alpha N_{c}}{\pi }}{\omega -\omega \left( a\right) -\omega \left( b\right)
-\omega \left( c\right) }  \notag \\
&&\times \frac{1}{\rho _{a}^{2}\rho _{b}^{2}\rho _{c}^{2}}\frac{b_{n,\nu
}b_{n_{a},\nu _{a}}b_{n_{b},\nu _{b}}b_{n_{c},\nu _{c}}\left\vert \rho \rho
_{a}\rho _{b}\rho _{c}\right\vert }{\left( 2\pi ^{2}\right) ^{4}}%
E_{Q}^{n,\nu }\left( \rho \right) E_{q_{a}}^{n_{a},\nu _{a}}\left( \rho
_{a}\right) E_{q_{b}}^{n_{b},\nu _{b}}\left( \rho _{b}\right)
E_{q_{c}}^{n_{c},\nu _{c}}\left( \rho _{c}\right)  \notag \\
&&\times g_{3P}\left( 1-h_{d},1-h_{a},1-h_{b}\right) g_{3P}\left(
1-h,h_{d},1-h_{c}\right)  \notag \\
&&\times \dint d^{2}\rho _{\alpha }d^{2}\rho _{\beta }d^{2}\rho _{\gamma
}d^{2}\rho _{\theta }d^{2}\rho _{\delta }\exp \left( iQ\rho _{\delta
}+iq_{c}\rho _{\theta }+iq_{a}\rho _{\alpha }+iq_{b}\rho _{\beta }\right)
\notag \\
&&\times \left( \rho _{\gamma \theta }^{-1+i^{\prime }}\rho _{\theta \delta
}^{-1+j^{\prime }}\rho _{\delta \gamma }^{-1+k^{\prime }}\right) \left(
\overline{\rho }_{\gamma \theta }^{-1+\overline{i}^{\prime }}\overline{\rho }%
_{\theta \delta }^{-1+\overline{j}^{\prime }}\overline{\rho }_{\delta \gamma
}^{-1+\overline{k}^{\prime }}\right)  \notag \\
&&\times \left( \rho _{\alpha \beta }^{-1+i^{\prime \prime }}\rho _{\beta
\gamma }^{-1+j^{\prime \prime }}\rho _{\gamma \alpha }^{-1+k^{\prime \prime
}}\right) \left( \overline{\rho }_{\alpha \beta }^{-1+\overline{i}^{\prime
\prime }}\overline{\rho }_{\beta \gamma }^{-1+\overline{j}^{\prime \prime }}%
\overline{\rho }_{\gamma \alpha }^{-1+\overline{k}^{\prime \prime }}\right)
\label{n3q}
\end{eqnarray}%
Hereafter in the computation of $n_{Y}^{(3)}$, we will restrict ourselves to
the limit of $\rho ^{2}Q^{2}\ll 1$ $\rho ^{2}q_{a}^{2}\ll 1$, $n_{Y}^{2}$, $%
\rho ^{2}q_{b}^{2}\ll 1$ and $\rho ^{2}q_{c}^{2}\ll 1$, where $Q^{2}$, $%
q_{a}^{2}$, $q_{b}^{2}$ and $q_{c}^{2}$ are of the same order. After
changing of variables,
\begin{eqnarray}
u &=&\frac{1}{2}\left( \rho _{\alpha }+\rho _{\beta }\right) +\frac{1}{2}%
\left( \rho _{\delta }+\rho _{\theta }\right) ,  \notag \\
v &=&\rho _{\alpha }-\rho _{\beta },  \notag \\
w &=&\rho _{\alpha }-\rho _{\gamma }+\rho _{\beta }-\rho _{\gamma },  \notag
\\
x &=&\rho _{\theta }-\rho _{\delta },  \notag \\
y &=&\rho _{\theta }-\rho _{\gamma }+\rho _{\delta }-\rho _{\gamma }
\label{cv2}
\end{eqnarray}%
the last three lines of Eq.(\ref{n3q}) can be written as%
\begin{eqnarray}
J &=&\frac{\pi ^{2}}{4}\delta ^{\left( 2\right) }\left(
Q+q_{a}+q_{b}+q_{c}\right)  \notag \\
&&\times \dint d^{2}vd^{2}w\exp \left[ iq_{ab}\frac{v}{2}+iq_{-}^{\prime }w%
\right] \left[ v^{-1+i^{\prime \prime }}\left( \frac{w-v}{2}\right)
^{-1+j^{\prime \prime }}\left( \frac{w+v}{2}\right) ^{-1+k^{\prime \prime }}%
\right] \times \left[ a.h.\right]  \notag \\
&&\times \int d^{2}xd^{2}y\exp \left[ iq_{cQ}\frac{x}{2}-iq_{-}^{\prime }y%
\right] \left[ x^{-1+j^{\prime }}\left( \frac{y-x}{2}\right) ^{-1+k^{\prime
}}\left( \frac{x+y}{2}\right) ^{-1+i^{\prime }}\right] \times \left[ a.h.%
\right] .
\end{eqnarray}%
where $a.h.$ stands for the anti-holomorphic part of the square-bracket
term, $q_{ab}=q_{a}-q_{b}$, $q_{cQ}=q_{c}-Q$ and $q_{-}^{\prime }=\frac{%
q_{a}+q_{b}-q_{c}-Q}{4}$. Following the same philosophy that we have
employed in the last section, we can expand $\exp \left[ iq_{-}^{\prime
}w-iq_{-}^{\prime }y\right] $ into Taylor series $\left( 1+iq_{-}^{\prime
}w+\cdots \right) \left( 1-iq_{-}^{\prime }y+\cdots \right) $, and calculate
the integrals order by order. One can easily see that all higher terms yield
the same anomalous dimension as the first term does(See appendix.\ref%
{general}). Thus, hereafter, we keep only the first term of the expansion%
\begin{eqnarray}
J_{0} &=&\frac{\pi ^{2}}{4}\delta ^{\left( 2\right) }\left(
Q+q_{a}+q_{b}+q_{c}\right)  \notag \\
&&\times \dint d^{2}vd^{2}w\exp \left[ i\left( q_{a}-q_{b}\right) \frac{v}{2}%
\right] \left[ v^{-1+i^{\prime \prime }}\left( \frac{w-v}{2}\right)
^{-1+j^{\prime \prime }}\left( \frac{w+v}{2}\right) ^{-1+k^{\prime \prime }}%
\right] \times \left[ a.h.\right]  \notag \\
&&\times \int d^{2}xd^{2}y\exp \left[ i\left( q_{c}-Q\right) \frac{x}{2}%
\right] \left[ x^{-1+j^{\prime }}\left( \frac{y-x}{2}\right) ^{-1+k^{\prime
}}\left( \frac{x+y}{2}\right) ^{-1+i^{\prime }}\right] \times \left[ a.h.%
\right] ,
\end{eqnarray}%
and then perform the integrals of coordinates
\begin{eqnarray}
J_{0} &=&\frac{\pi ^{2}}{4}\delta ^{\left( 2\right) }\left(
Q+q_{a}+q_{b}+q_{c}\right)  \notag \\
&&\times 4\pi ^{2}\left( \frac{2}{q_{ab}}\right) ^{i^{\prime \prime
}+j^{\prime \prime }+k^{\prime \prime }+\overline{i}^{\prime \prime }+%
\overline{j}^{\prime \prime }+\overline{k}-2}\frac{\Gamma \left( k^{\prime
\prime }\right) \Gamma \left( j^{\prime \prime }\right) }{\Gamma \left(
k^{\prime \prime }+j^{\prime \prime }\right) }\frac{\Gamma \left( 1-%
\overline{j}^{\prime \prime }-\overline{k}^{\prime \prime }\right) }{\Gamma
\left( 1-\overline{j}^{\prime \prime }\right) \Gamma \left( 1-\overline{k}%
^{\prime \prime }\right) }\frac{\Gamma \left( i^{\prime \prime }+j^{\prime
\prime }+k^{\prime \prime }-1\right) }{\Gamma \left( 2-\overline{i}^{\prime
\prime }-\overline{j}^{\prime \prime }-\overline{k}^{\prime \prime }\right) }
\notag \\
&&\times 4\pi ^{2}\left( \frac{2}{q_{cQ}}\right) ^{i^{\prime }+j^{\prime
}+k^{\prime }+\overline{i}^{\prime }+\overline{j}+\overline{k}^{\prime }-2}%
\frac{\Gamma \left( k^{\prime }\right) \Gamma \left( i^{\prime }\right) }{%
\Gamma \left( k^{\prime }+i^{\prime }\right) }\frac{\Gamma \left( 1-%
\overline{i}^{\prime }-\overline{k}^{\prime }\right) }{\Gamma \left( 1-%
\overline{i}^{\prime }\right) \Gamma \left( 1-\overline{k}^{\prime }\right) }%
\frac{\Gamma \left( i^{\prime }+j^{\prime }+k^{\prime }-1\right) }{\Gamma
\left( 2-\overline{i}^{\prime }-\overline{j}^{\prime }-\overline{k}^{\prime
}\right) }.
\end{eqnarray}%
In addition, at leading order in rapidity, one should pick the $\frac{1}{%
\omega -\omega \left( a\right) -\omega \left( b\right) -\omega \left(
c\right) }$ pole in $\dint \frac{d\omega }{2\pi i}$ integral, and use saddle
point approximation to evaluate $\dint dh_{a}dh_{b}dh_{c}$ integrals which
eventually fixes $h_{a}=h_{b}=h_{c}=\frac{1}{2}$, $n_{a}=n_{b}=n_{c}=0$.
Moreover, we assume $n=n_{d}=0$ in the following discussion. (In $%
n_{Y}^{(2)} $ calculation, we have shown that higher $n$ is suppressed. Here
it is our assumption that higher $n$ and $n_{d}$ would also be suppressed.
Nevertheless, we have been unable to find a proof for this point since the
calculation becomes very lengthy.) Thus,
\begin{eqnarray}
n_{Y,Q,q_{a},q_{b},q_{c}}^{(3)\rho ,\rho _{a},\rho _{b},\rho _{c}} &\simeq &%
\frac{\pi ^{2}}{4}\delta ^{\left( 2\right) }\left(
Q+q_{a}+q_{b}+q_{c}\right) \dint dhdh_{d}\frac{\frac{1}{2}}{3\chi \left(
\frac{1}{2}\right) -\chi \left( \gamma \right) }\frac{\frac{1}{2}}{2\chi
\left( \frac{1}{2}\right) -\chi \left( \gamma _{d}\right) }  \notag \\
&&\times \frac{b_{0,\nu }\left\vert \rho \right\vert }{2\pi ^{2}}%
E_{Q}^{0,\nu }\left( \rho \right) g_{3P}\left( 1-\gamma _{d},\frac{1}{2},%
\frac{1}{2}\right) g_{3P}\left( 1-\gamma ,\gamma _{d},\frac{1}{2}\right)
\notag \\
&&\times 4\pi ^{2}\left( \frac{4}{q_{ab}^{2}}\right) ^{\gamma _{d}-1}\frac{%
\Gamma ^{2}\left( \gamma _{d}\right) \Gamma \left( \frac{1}{2}-\gamma
_{d}\right) 2^{1-4\gamma _{d}}}{\Gamma ^{2}\left( 1-\gamma _{d}\right)
\Gamma \left( \frac{1}{2}+\gamma _{d}\right) }  \notag \\
&&\times 4\pi ^{2}\left( \frac{4}{q_{cQ}^{2}}\right) ^{\gamma -\gamma _{d}}%
\frac{\Gamma ^{2}\left( -\gamma _{d}+\frac{1}{2}+\gamma \right) \Gamma
\left( \frac{3}{2}-\gamma _{d}-\gamma \right) \Gamma \left( 2\gamma
_{d}-1\right) }{\Gamma ^{2}\left( \gamma _{d}+\frac{1}{2}-\gamma \right)
\Gamma \left( \frac{1}{2}+\gamma _{d}+\gamma \right) \Gamma \left( 2-2\gamma
_{d}\right) }  \notag \\
&&\times \frac{1}{8\pi ^{3}}\frac{e\left( \rho _{a},q_{a}\right) e\left(
\rho _{b},q_{b}\right) e\left( \rho _{c},q_{c}\right) }{\rho _{a}q_{ab}\rho
_{b}q_{ab}\rho _{c}q_{cQ}}\frac{\exp \left[ 3\left( \alpha _{P}-1\right) Y%
\right] \exp \left[ -\frac{\pi \left( \ln ^{2}\left( q_{a}\rho _{a}\right)
+\ln ^{2}\left( q_{b}\rho _{b}\right) +\ln ^{2}\left( q_{c}\rho _{c}\right)
\right) }{28\alpha C_{F}\zeta \left( 3\right) Y}\right] }{\left( 7\alpha
\zeta \left( 3\right) C_{F}Y\right) ^{9/2}}
\end{eqnarray}%
Unfortunately, we are unable to to evaluate $\dint dh_{d}$ since $\frac{%
q_{ab}^{2}}{q_{cQ}^{2}}$ is not necessarily a large parameter. This
difficulty originates from the fact that all the relevant length scales have
been integrated out. In this case we have to take into account all the
singularities according to the residue theorem, or we can just choose the
vertical contour from $\frac{1}{2}-i\infty $ to $\frac{1}{2}+i\infty $ and
define the integral as a function of $\gamma $. Let us define
\begin{eqnarray}
dh_{d}-integral &\simeq &\int_{\frac{1}{2}-i\infty }^{\frac{1}{2}+i\infty }%
\frac{d\gamma _{d}}{2\pi i}\frac{1}{\pi ^{3}}\frac{1}{2\chi \left( \frac{1}{2%
}\right) -\chi \left( \gamma _{d}\right) }g_{3P}\left( 1-\gamma _{d},\frac{1%
}{2},\frac{1}{2}\right) g_{3P}\left( 1-\gamma ,\gamma _{d},\frac{1}{2}\right)
\notag \\
&&\times \left( 4\pi ^{2}\right) ^{2}\left( \frac{q_{ab}^{2}}{q_{cQ}^{2}}%
\right) ^{1-\gamma _{d}}\frac{\Gamma ^{3}\left( \gamma _{d}\right) \Gamma
^{2}\left( -\gamma _{d}+\frac{1}{2}+\gamma \right) \Gamma \left( \frac{3}{2}%
-\gamma _{d}-\gamma \right) }{\Gamma ^{3}\left( 1-\gamma _{d}\right) \Gamma
^{2}\left( \gamma _{d}+\frac{1}{2}-\gamma \right) \Gamma \left( \frac{1}{2}%
+\gamma _{d}+\gamma \right) } \\
&=&16\pi \frac{q_{ab}^{2}}{q_{cQ}^{2}}f\left( \gamma \right) .
\end{eqnarray}%
Here one can not use the residue theorem since the ratio $\frac{q_{ab}^{2}}{%
q_{cQ}^{2}}$ is not necessarily large or small. Nevertheless, it is
straightforward to see that $f\left( \gamma \right) $ should be analytic in
the strip domain $0<\func{Re}\gamma <1$ when $\gamma _{d}$ is integrated
from $\frac{1}{2}-i\infty $ to $\frac{1}{2}+i\infty $. Thus,
\begin{eqnarray}
n_{Y,Q,q_{a},q_{b},q_{c}}^{(3)\rho ,\rho _{a},\rho _{b},\rho _{c}} &\simeq
&\delta ^{\left( 2\right) }\left( Q+q_{a}+q_{b}+q_{c}\right) \dint dh\frac{1%
}{3\chi \left( \frac{1}{2}\right) -\chi \left( \gamma \right) }\frac{%
b_{0,\nu }\left\vert \rho \right\vert }{2\pi ^{2}}E_{Q}^{0,\nu }\left( \rho
\right) f\left( \gamma \right) \left( \frac{4}{q_{cQ}^{2}}\right) ^{\gamma
-1}  \notag \\
&&\times \frac{1}{8}\frac{e\left( \rho _{a},q_{a}\right) e\left( \rho
_{b},q_{b}\right) e\left( \rho _{c},q_{c}\right) }{\rho _{a}q_{cQ}\rho
_{b}q_{cQ}\rho _{c}q_{cQ}}\frac{\exp \left[ 3\left( \alpha _{P}-1\right) Y%
\right] \exp \left[ -\frac{\pi \left( \ln ^{2}\left( q_{a}\rho _{a}\right)
+\ln ^{2}\left( q_{b}\rho _{b}\right) +\ln ^{2}\left( q_{c}\rho _{c}\right)
\right) }{28\alpha C_{F}\zeta \left( 3\right) Y}\right] }{\left( 7\alpha
\zeta \left( 3\right) C_{F}Y\right) ^{9/2}}.
\end{eqnarray}%
One can cast the final integral into
\begin{eqnarray}
n_{Y,Q,q_{a},q_{b},q_{c}}^{(3)\rho ,\rho _{a},\rho _{b},\rho _{c}} &\simeq
&\delta ^{\left( 2\right) }\left( Q+q_{a}+q_{b}+q_{c}\right) \dint \frac{%
d\gamma }{2\pi i}\frac{f\left( \gamma \right) }{3\chi \left( \frac{1}{2}%
\right) -\chi \left( \gamma \right) }\frac{2^{4\gamma }}{2\pi ^{2}}\frac{%
\Gamma \left( 1-\gamma \right) \Gamma \left( \gamma +\frac{1}{2}\right) }{%
\Gamma \left( \gamma \right) \Gamma \left( \frac{1}{2}-\gamma \right) }%
\left\vert \rho \right\vert E_{Q}^{0,\nu }\left( \rho \right) \left( \frac{4%
}{q_{cQ}^{2}}\right) ^{\gamma -1}  \notag \\
&&\times \frac{1}{8}\frac{e\left( \rho _{a},q_{a}\right) e\left( \rho
_{b},q_{b}\right) e\left( \rho _{c},q_{c}\right) }{\rho _{a}q_{cQ}\rho
_{b}q_{cQ}\rho _{c}q_{cQ}}\frac{\exp \left[ 3\left( \alpha _{P}-1\right) Y%
\right] \exp \left[ -\frac{\pi \left( \ln ^{2}\left( q_{a}\rho _{a}\right)
+\ln ^{2}\left( q_{b}\rho _{b}\right) +\ln ^{2}\left( q_{c}\rho _{c}\right)
\right) }{28\alpha C_{F}\zeta \left( 3\right) Y}\right] }{\left( 7\alpha
\zeta \left( 3\right) C_{F}Y\right) ^{9/2}}.
\end{eqnarray}%
In the end, following the same procedure as in the last section, we write $%
E_{Q}^{0,\nu }\left( \rho \right) \simeq \rho ^{1-2\gamma }-\frac{Q^{2\gamma
-1}\Gamma ^{2}\left( \frac{3}{2}-\gamma \right) 2^{6-12\gamma }}{\Gamma
^{2}\left( \frac{1}{2}+\gamma \right) }\left( \rho Q\right) ^{2\gamma -1}$
in $\rho Q\ll 1$ limit\cite{Navelet:1997xn}, close the contour to the left
for the first term, and close the contour to the right for the second.
Therefore, according to the residue theorem, the contour integral equals to
the sum of all residues enclosed by the contour. The dominant contribution
is
\begin{eqnarray}
n_{Y,Q,q_{a},q_{b},q_{c}}^{(3)\rho ,\rho _{a},\rho _{b},\rho _{c}} &\propto &%
\frac{\left( \rho ^{2}q_{cQ}^{2}\right) ^{1-\gamma _{0}^{\left( 3\right)
\ast }}}{\rho _{a}q_{cQ}\rho _{b}q_{cQ}\rho _{c}q_{cQ}}\frac{\exp \left[
3\left( \alpha _{P}-1\right) Y\right] \exp \left[ -\frac{\pi \left( \ln
^{2}\left( q_{a}\rho _{a}\right) +\ln ^{2}\left( q_{b}\rho _{b}\right) +\ln
^{2}\left( q_{c}\rho _{c}\right) \right) }{28\alpha C_{F}\zeta \left(
3\right) Y}\right] }{\left( 7\alpha \zeta \left( 3\right) C_{F}Y\right)
^{9/2}}  \notag \\
&&\times \delta ^{\left( 2\right) }\left( Q+q_{a}+q_{b}+q_{c}\right) e\left(
\rho _{a},q_{a}\right) e\left( \rho _{b},q_{b}\right) e\left( \rho
_{c},q_{c}\right)  \label{n3f}
\end{eqnarray}%
where $\gamma _{0}^{\left( 3\right) \ast }=0.12$ and $1-\gamma _{0}^{\left(
3\right) \ast }=0.88$ are the two solutions of the equation $3\chi \left(
\frac{1}{2}\right) -\chi \left( \gamma \right) =0$ in the domain $\left[ 0,1%
\right] $. The anomalous dimension $\gamma _{0}^{\left( 3\right) \ast }=0.12$
found here is new.

In the case of onium-onium scattering, the triple pomeron exchange amplitude
between two onia with sizes $\rho $ and $\rho ^{\prime }$ then scales as
\begin{eqnarray}
F^{\left( 3\right) }\left( \rho ,\rho ^{\prime },q,Y\right) \propto \left(
q^{2}\rho ^{2}\right) ^{0.88}\left( q^{2}\rho ^{\prime 2}\right)
^{0.88}\left\{ \alpha ^{2}\frac{\exp \left[ \left( \alpha _{P}-1\right) Y%
\right] }{\left( 7\alpha \zeta \left( 3\right) C_{F}Y\right) ^{3/2}}\right\}
^{3}.  \label{f3}
\end{eqnarray}

\section{Generalization to $1\Rightarrow k$ amplitude}

\begin{figure}[tbp]
\begin{center}
\includegraphics[width=6.5cm]{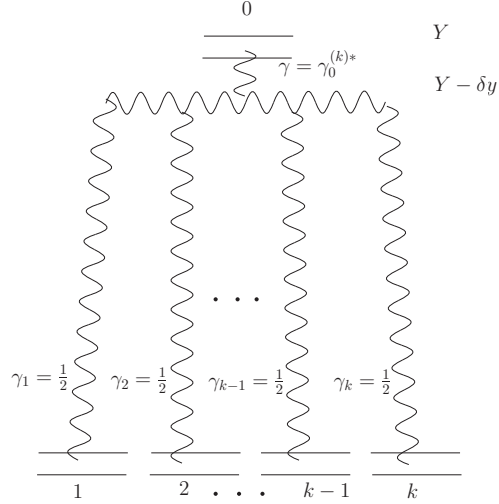}
\end{center}
\caption[*]{Assuming $\protect\delta y$ is infinitesimal, the anomalous
dimensions of $1\Rightarrow k$ amplitude $n_{Y}^{k}$ are illustrated in this
figure.}
\label{nk1}
\end{figure}
Based on the above computation, one can easily conjecture that the anomalous
dimensions of $1\Rightarrow k$ amplitudes should be governed by the equation
$k\chi \left( \frac{1}{2}\right) -\chi \left( \gamma \right) =0$, which
comes from the dynamical pole introduced by the rapidity integration.
Another way of stating this conjecture is that $1\Rightarrow k$ amplitudes
tend to have constant energy dependence from $0$ to $Y$(see Fig.~\ref{nk1})
as a result of $k\chi \left( \frac{1}{2}\right) =\chi \left( \gamma \right) $%
. Namely, the total BFKL intercepts of multiple pomeron exchanges are
independent of rapidity $y$. Supposing $\gamma _{0}^{\left( k\right) \ast }$
is the solution to this equation between $0$ and $\frac{1}{2}$, then the $%
1\Rightarrow k$ amplitude should scale as
\begin{eqnarray}
n_{Y\left( q,q_{1}\cdots q_{k}\right) }^{\left( k\right) \left( \rho ,\rho
_{1}\cdots \rho _{k}\right) } &\varpropto &\frac{\left( \rho
^{2}q^{2}\right) ^{1-\gamma _{0}^{\left( k\right) \ast }}}{\left( \rho
_{1}q\right) \cdots \left( \rho _{k}q\right) }\frac{\exp \left[ k\left(
\alpha _{P}-1\right) Y\right] \exp \left[ -\frac{\pi \left[ \ln ^{2}\left(
\rho q\right) +\ln ^{2}\left( \rho _{1}q_{1}\right) +\cdots +\ln ^{2}\left(
\rho _{k}q_{k}\right) \right] }{28\alpha C_{F}\zeta \left( 3\right) Y}\right]
}{\left( 7\alpha \zeta \left( 3\right) C_{F}Y\right) ^{3k/2}} \\
&&\times \delta ^{\left( 2\right) }\left( q+q_{1}+\cdots +q_{k}\right)
e\left( \rho _{1},q_{1}\right) \cdots e\left( \rho _{k},q_{k}\right) .
\end{eqnarray}%
For reader's convenience, we list the first five $\gamma _{0}^{\left(
k\right) \ast }$ in the following: $\gamma _{0}^{\left( 1\right) \ast }=%
\frac{1}{2}$, $\gamma _{0}^{\left( 2\right) \ast }=0.18$, $\gamma
_{0}^{\left( 3\right) \ast }=0.12$, $\gamma _{0}^{\left( 4\right) \ast
}=0.090$, $\gamma _{0}^{\left( 5\right) \ast }=0.072$. This result fits the
naive expectation that the anomalous dimension should approach zero in small
rapidity limit or in DGLAP limit. Furthermore, when $k$ becomes large
enough, we obtain $\gamma _{0}^{\left( k\right) \ast }\simeq 0$. This
indicates that the anomalous dimensions of large order pomeron loops are
dominated by the DGLAP evolution. We should remind the reader that this
conjecture can only be used in the region $\ln ^{2}\left( q\rho \right) \ll
\frac{14\alpha C_{F}\zeta \left( 3\right) }{\pi }Y$ and $\rho ^{2}q^{2}\ll 1$%
, while $q,q_{1}\cdots , q_{k}$ are all of the same order, and so are all
the dipole sizes $\rho ,\rho _{1}\cdots ,\rho _{k}$. In the case of
onium-onium scattering, we obtain that the corresponding k-pomeron exchange
amplitude should scale as
\begin{equation}
F^{\left( k\right) }\left( \rho ,\rho ^{\prime },q,Y\right) \propto \left(
q^{2}\rho ^{2}\right) ^{1-\gamma _{0}^{\left( k\right) \ast }}\left(
q^{2}\rho ^{\prime 2}\right) ^{1-\gamma _{0}^{\left( k\right) \ast }}\left\{
\alpha ^{2}\frac{\exp \left[ \left( \alpha _{P}-1\right) Y\right] }{\left(
7\alpha \zeta \left( 3\right) C_{F}Y\right) ^{3/2}}\right\} ^{k},
\end{equation}%
where we restrict ourselves to the $\rho ^{2}q^{2}\ll 1$ and $\rho ^{\prime
2}q^{2}\ll 1$ limit. As a result, different order of pomeron-loop amplitudes
have different dipole size dependence, namely, they belong to different
universality classes. Thus, it seems that re-summation of leading order
amplitudes is insufficient and impossible to reach unitarity. The argument
is straightforward: supposing that unitarity is achieved by resummation of
all loop amplitudes by means of delicate balances for fixed $\rho $, it
seems impossible to obtain unitarity again when $\rho $ is changed to
another value since the delicate balance is then broken.

In summary, according to the conjectured formula, we have found infinite
number of new anomalous dimensions between $\frac{1}{2}$ and $0$. Comparing
the multiple pomeron exchanges to single pomeron exchange, it seems that the
evolution gets pushed towards DGLAP evolution\cite{Gribov:1972ri}, and the
anomalous dimension discretely approaches the anomalous dimension of DGLAP
evolution. Because of the lack of the rapidity space for upper and lower
pomerons, the evolutions in pomeron loops manage to balance themselves in
between the BFKL evolution and the DGLAP evolution. This explains why the
new anomalous dimensions are distributed in between $\frac{1}{2}$ and $0$.
Furthermore, it is intuitive to notice that the saturation anomalous
dimension found in ref.~\cite{Mueller:2002zm} is $\gamma_{s} = 0.37$ as a
solution to Kovchegov equation in geometric scaling region. Kovchegov
equation essentially resums multiple pomeron exchanges (fan diagrams) and
yields an anomalous dimension between $\frac{1}{2}$ and $0$ as a result of
re-summation of multiple pomeron exchanges. The explicit connection between
the $\gamma_{s}$ and $\gamma _{0}^{\left( k\right) \ast }$, however, is
still unknown and remains as an open question.

\section{Comparison with reggeon field theory calculus}

Reggeon field theory (RFT) calculus(for a review, see ref.~\cite%
{Baker:1976cv,Collins:1977jy,Forshaw:1997dc,Donnachie:2002en}), similar as
Feynman rules, provides definite intercepts ($\alpha \left(t\right)-1=\alpha
\left(0\right) +\alpha ^{\prime} t$) and propagators for reggeons (including
pomerons) in QCD phenomenology. In RFT, the pomeron anomalous dimension $%
\gamma =\frac{1}{2}$ is a universal (conserved) quantity and it corresponds
to the pomeron intercept $\alpha _{P}-1=\frac{8\alpha C_{F}\ln 2}{\pi }$. In
leading order of $1\Rightarrow 2$ amplitude, RFT has a genuine triple
pomeron vertex which connects 3 pomerons with anomalous dimension $\gamma =%
\frac{1}{2}$. This approach certainly is justified in some situations such
as large diffractive mass scattering, where the large diffractive mass is
large enough to fix the anomalous dimension of upper pomeron at $\frac{1}{2}$
by saddle point approximation(e.g., see ref.~\cite{Bialas:1997xp}). On the
other hand, as far as the unitarity problem is concerned, one has to
integrate the intermediate rapidity to the upper limit $Y$, in which case
the anomalous dimension of the upper pomeron in the vertex is no longer
fixed at $\frac{1}{2}$.

Following this philosophy and the essence of the RFT, in order to compare
with the calculation we have finished above, let us re-consider $n_{Y}^{(2)}$
in forward scattering ($Q=0,q_{a}=-q_{b}=q$) by assuming that $Y-y$ is
always large enough to justify the saddle point approximation (In fact, this
assumption breaks down when one integrates rapidity to the upper limit $Y$.)
First of all, let us begin with the expression with the rapidity
integration,
\begin{eqnarray}
n_{Y,Q,q_{-},0}^{(2)\rho ,\rho _{a},\rho _{b}} &=&\frac{1}{\rho _{a}^{2}\rho
_{b}^{2}}\int_{0}^{Y}\frac{\alpha N_{c}}{\pi }dy\dint dh\exp \left[ \omega
\left( \gamma \right) \left( Y-y\right) \right]  \notag \\
&&\times \frac{b_{n,\nu }\left\vert \rho \right\vert }{2\pi ^{2}}%
E_{0}^{n,\nu }\left( \rho \right) g_{3P}\left( 1-h,\frac{1}{2},\frac{1}{2}%
\right) 4\pi ^{2}\left( \frac{4}{q^{2}}\right) ^{\gamma -1}\frac{\Gamma
^{3}\left( \gamma \right) \Gamma \left( 1-2\gamma \right) }{\Gamma \left(
2\gamma \right) \Gamma ^{3}\left( 1-\gamma \right) }  \notag \\
&&\times \frac{1}{4\pi ^{2}}\frac{\rho _{a}\rho _{b}}{q^{2}}e\left( \rho
_{a},q\right) e\left( \rho _{b},-q\right) \frac{\exp \left[ 2\left( \alpha
_{P}-1\right) y\right] \exp \left[ -\frac{\pi \left( \ln ^{2}\left( q\rho
_{a}\right) +\ln ^{2}\left( q\rho _{b}\right) \right) }{28\alpha C_{F}\zeta
\left( 3\right) y}\right] }{\left( 7\alpha \zeta \left( 3\right)
C_{F}y\right) ^{3}}
\end{eqnarray}%
Dropping all the higher $n$, and changing $\dint dh$ into $\dint \frac{%
d\gamma }{2\pi i}$, one can simplify $n_{Y}^{(2)}$ and get,
\begin{eqnarray}
n_{y,Y,Q,q_{-},0}^{(2)\rho ,\rho _{a},\rho _{b}} &\simeq &\dint \frac{%
d\gamma }{2\pi i}\left( \frac{q^{2}\rho ^{2}}{4}\right) ^{1-\gamma }\exp %
\left[ \omega \left( \gamma \right) \left( Y-y\right) \right] \frac{4\Gamma
\left( \gamma \right) }{\Gamma \left( 1-\gamma \right) }g_{3P}\left(
1-\gamma ,\frac{1}{2},\frac{1}{2}\right)  \notag \\
&&\times \frac{1}{4\pi ^{2}}\frac{1}{q^{2}\rho _{a}\rho _{b}}e\left( \rho
_{a},q\right) e\left( \rho _{b},-q\right) \frac{\exp \left[ 2\left( \alpha
_{P}-1\right) y\right] \exp \left[ -\frac{\pi \left( \ln ^{2}\left( q\rho
_{a}\right) +\ln ^{2}\left( q\rho _{b}\right) \right) }{28\alpha C_{F}\zeta
\left( 3\right) y}\right] }{\left( 7\alpha \zeta \left( 3\right)
C_{F}y\right) ^{3}}
\end{eqnarray}%
If one uses saddle point approximation to evaluate the $\dint \frac{d\gamma
}{2\pi i}$ integral before the rapidity integral, one reaches,
\begin{eqnarray}
n_{y,Y,Q,q_{-},0}^{(2)\rho ,\rho _{a},\rho _{b}} &\simeq &\frac{1}{8\pi ^{2}}%
\frac{q\rho }{q\rho _{a}q\rho _{b}}g_{3P}\left( \frac{1}{2},\frac{1}{2},%
\frac{1}{2}\right) e\left( \rho _{a},q\right) e\left( \rho _{b},-q\right)
\notag \\
&&\times \frac{\exp \left[ 2\left( \alpha _{P}-1\right) y+\left( \alpha
_{P}-1\right) \left( Y-y\right) \right] \exp \left[ -\frac{\pi \left( \ln
^{2}\left( q\rho _{a}\right) +\ln ^{2}\left( q\rho _{b}\right) \right) }{%
28\alpha C_{F}\zeta \left( 3\right) y}-\frac{\pi \left( \ln ^{2}\left( q\rho
\right) \right) }{28\alpha C_{F}\zeta \left( 3\right) \left( Y-y\right) }%
\right] }{\left( 7\alpha \zeta \left( 3\right) C_{F}y\right) ^{3}\left[
7\alpha \zeta \left( 3\right) C_{F}\left( Y-y\right) \right] ^{1/2}}
\end{eqnarray}%
Indeed, we now obtain the genuine triple pomeron vertex which connects 3
pomerons with anomalous dimension $\gamma =\frac{1}{2}$. Finishing the
rapidity integration yields,
\begin{eqnarray}
n_{Y,Q,q_{-},0}^{(2)\rho ,\rho _{a},\rho _{b}} &\simeq &\frac{1}{8\pi ^{2}}%
\frac{q\rho }{q\rho _{a}q\rho _{b}}g_{3P}\left( \frac{1}{2},\frac{1}{2},%
\frac{1}{2}\right) e\left( \rho _{a},q\right) e\left( \rho _{b},-q\right) \\
&&\times \sqrt{\frac{1}{14\zeta \left( 3\right) \ln 2}}\frac{\exp \left[
2\left( \alpha _{P}-1\right) Y\right] \exp \left[ -\frac{\pi \left( \ln
^{2}\left( q\rho _{a}\right) +\ln ^{2}\left( q\rho _{b}\right) \right) }{%
28\alpha C_{F}\zeta \left( 3\right) Y}\right] }{\left( 7\alpha \zeta \left(
3\right) C_{F}Y\right) ^{3}}
\end{eqnarray}

This result agrees with eq.(56) in ref.\cite{Mueller:1994jq}. Integrating
the triple pomeron vertex found in eq.(56) in ref.\cite{Mueller:1994jq}, one
gets
\begin{eqnarray}
\int_{0}^{Y}d\overline{y}\overline{n}_{2}\left( Y,\overline{y}%
,x_{01},q\right) &=&\int_{0}^{Y}d\overline{y}\frac{\alpha
C_{F}V_{0}x_{01}\exp \left[ 2\left( \alpha _{P}-1\right) \overline{y}+\left(
\alpha _{P}-1\right) \left( Y-\overline{y}\right) \right] }{8q\sqrt{7\alpha
\zeta \left( 3\right) C_{F}\left( Y-\overline{y}\right) }\left( 7\alpha
\zeta \left( 3\right) C_{F}\overline{y}\right) ^{3}} \\
&\simeq &\frac{\pi V_{0}\exp \left[ 2\left( \alpha _{P}-1\right) Y\right] }{%
16\sqrt{14\zeta \left( 3\right) \ln 2}\left( 7\alpha \zeta \left( 3\right)
C_{F}Y\right) ^{3}}\frac{x_{01}}{q}
\end{eqnarray}%
It clearly differs from Eq.(\ref{n2}) which is proportional to $\frac{\left(
\rho ^{2}q_{-}^{2}\right) ^{1-\gamma _{0}^{\left( 2\right) \ast }}}{\rho
_{a}q_{-}\rho _{b}q_{-}}g_{3P}\left( 0.82,\frac{1}{2},\frac{1}{2}\right) $
as a result of the anomalous dimension being $1-\gamma _{0}^{\left( 2\right)
\ast }$. The origin of this difference comes from the rapidity integral
which dynamically changes the anomalous dimension from $\frac{1}{2}$ to $%
\gamma _{0}^{\left( 2\right) \ast }$.

The zero transverse dimension toy models\cite%
{Rembiesa:2005gj,Shoshi:2005pf,Shoshi:2006eb,Bondarenko:2006rh,Kozlov:2006zj,Blaizot:2006wp}%
, which catch the essence of reggeon field theory calculus, also have
universal rules for the pomeron intercepts and propagators, as well as the
triple pomeron (or reggeon) vertices. The toy model does not contain the
anomalous dimension or transverse dimensions. In some sense, it
over-simplifies the problem and fails to catch new features of the QCD
pomeron that we found above.

On the other hand, the QCD dipole model, as one can easily spot from the
calculation in Sections \ref{single}, \ref{double2} and \ref{triple3},
contains the distinct feature of the non-existence of universal intercepts
and anomalous dimensions for QCD pomerons in the unitarity calculation. In
leading order calculation of $1\Rightarrow k$ amplitude, we always push the
rapidity integral to the upper limit $Y$ which leaves infinitesimal rapidity
for the upper pomeron (see Fig. \ref{nk1}). As a result, the anomalous
dimensions of this pomeron are then fixed by the dynamical pole introduced
by the rapidity integration. The anomalous dimensions are no longer
universal and they vary from graph to graph according to the detail
structure of the graph and dynamics. Nevertheless, the other side of the
coin is that we now have the constant energy dependence (coefficients of the
rapidity) from $0$ to $Y$ while this certainly is not true in RFT.

\section{Conclusion}

We explicitly calculate the anomalous dimensions of the single, double
pomeron and triple pomeron exchange amplitudes in QCD dipole model in the
leading logarithmic approximation. Other than the usual pomeron anomalous
dimension $\frac{1}{2}$, we find $\gamma _{0}^{(2)\ast }=0.18$ and $\gamma
_{0}^{(3)\ast }=0.12$ for double and triple pomeron exchange amplitude,
respectively. Based on the calculation, the general formula $\gamma
_{0}^{(k)\ast }=\chi ^{-1}\left( k\chi \left( \frac{1}{2}\right) \right) $
which governs anomalous dimensions of $1\Rightarrow k$ amplitude is
conjectured in the region $\ln ^{2}\left( q\rho \right) \ll \frac{14\alpha
C_{F}\zeta \left( 3\right) }{\pi }Y$ and $\rho ^{2}q^{2}\ll 1$, where $\rho $
stands for all the dipole sizes and $q$ represents all the momenta scales.

Furthermore, the calculation of forward scattering $n_{Y}^{\left( 2\right) }$
in the $n\neq 0$ case shows that contributions of sub-dominant trajectories
of the upper pomeron are suppressed by powers of $\rho ^{2}q^{2}$ in $\rho
^{2}q^{2}\ll 1$ limit. It is our conjecture that sub-dominant trajectories
can be neglect in the non-forward scattering case. We utilize this
conjecture in the calculation of $n_{Y}^{\left( 3\right) }$ and get rid of
the contribution from nonzero value of $n_{d}$.

In addition, different pomeron loop amplitudes (leading order) belong to
different universality class as a result of different anomalous dimensions.
It seems that re-summation of these amplitudes is insufficient and
impossible to reach unitarity. One may have to take higher order
contributions into account.

Last but not least, in comparison with the reggeon field theory, one finds
that there are two differences between this computation and the reggeon
field theory although the BFKL characteristic function $\chi \left( \gamma
\right) $ is universal. The first one is that the anomalous dimensions are
no longer a constant and they vary according to their positions in the graph
while $\gamma =\frac{1}{2}$ everywhere in RFT in order to have a fixed
pomeron intercept(for small and fixed $t$ value); the second difference is
that the QCD dipole model tends to have a constant energy dependence
(coefficients of the rapidity) from $0$ to $Y$ while this certainly is not
true in RFT. Namely, the total BFKL intercepts of multiple pomeron exchanges
are independent of rapidity $y$ in QCD dipole model. The constancy of the
energy dependence is equivalent to the general formula $\gamma _{0}^{(k)\ast
}=\chi ^{-1}\left( k\chi \left( \frac{1}{2}\right) \right) $ and can easily
explain its physical meaning.

\begin{acknowledgments}
I am grateful to Professor A.H. Mueller for suggesting this work and
numerous inspiring discussions.

I acknowledge the helpful discussions with Edmond Iancu, Cyrille
Marquet, Stephane Munier, Arif Shoshi and Gregory Soyez, as well
as the hospitality and support of SPhT Saclay. I would like to
thank L. Motyka for communications as well.

I also wish to thank Fakult{\"{a}}t f{\"{u}}r Physik of Universit{\"{a}}t
Bielefeld, II. Institut f{\"{u}}r Theoretische Physik of Universit{\"{a}}t
Hamburg and DESY, as well as the Galileo Galilei Institute for Theoretical
Physics for the hospitality and the INFN for partial support during my visit
when this work was initialized.

\end{acknowledgments}

\appendix

\section{The evaluation of integrals.}

\label{integral}

In this and the next appendices, we evaluate the following integral,
\begin{equation}
I_{Q,q_{-}}^{h,h_{a},h_{b}}=\dint d^{2}vd^{2}w\exp \left( -iQ\cdot \frac{w}{2%
}+iq_{-}\cdot v\right) \left[ v^{-1+i}\left( \frac{w-v}{2}\right)
^{-1+j}\left( \frac{w+v}{2}\right) ^{-1+k}\right] \left[ a.h.\right] .
\end{equation}%
where a.h. stands for the antiholomorphic part of the square-bracket term.
First of all, let us perform the Taylor expansion of the $\exp \left( -iQ%
\frac{w}{2}\right) $, and keep only the first term $1$. We will discuss the
results for higher terms in the next appendix. Thus, the above integral
becomes,
\begin{eqnarray}
I_{Q,q_{-}}^{h,h_{a},h_{b},0}=\dint d^{2}vd^{2}w\exp \left( iq_{-}\cdot
v\right) \left[ v^{-1+i}\left( \frac{w-v}{2}\right) ^{-1+j}\left( \frac{w+v}{%
2}\right) ^{-1+k}\right] \left[ a.h.\right] .  \label{inte}
\end{eqnarray}

\subsection{The $d^{2}w$ integral}

Furthermore, one can decouple Eq.(\ref{inte}) into two 2-dim integrals in
which the $d^{2}w$ integral can be reduced to the following integral:
\begin{equation}
I=\dint \frac{dzd\overline{z}}{2i}z^{-1+i}\overline{z}^{-1+\overline{i}%
}\left( 1-z\right) ^{j-1}\left( 1-\overline{z}\right) ^{\overline{j}-1},
\end{equation}
where $i-\overline{i}$ and $j-\overline{j}$ are integers. The solution to
this integral can be found in Dotsenko and Fateev's paper\cite%
{Dotsenko:1984nm} in statistical physics. In the following, we carry out the
detailed calculation\cite{Munier} in complex plane.

The first step is to change $z=x+iy, \overline{z}=x-iy$, and then perform a
Wick rotation $y \rightarrow i\left( 1-2i\epsilon \right)y$, where the $%
\epsilon$, which is an infinitesimal positive number, makes sure that the
singularities are not touched. We obtain,

\begin{equation}
I=i\int_{-\infty }^{+\infty }dx\int_{-\infty }^{+\infty }dy\left(
x-y+2i\epsilon y\right) ^{i-1}\left( x+y-2i\epsilon y\right) ^{\overline{i}%
-1}\left( 1-x+y-2i\epsilon y\right) ^{j-1}\left( 1-x-y+2i\epsilon y\right) ^{%
\overline{j}-1}.
\end{equation}

Next, one can change the variables into $X_{+}=x+y$ and $X_{-}=x-y$ and cast
the integral into the form,

\begin{figure}[tbp]
\begin{center}
\includegraphics[width=5cm]{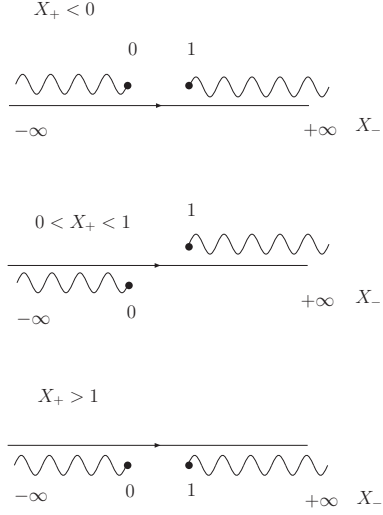} 
\end{center}
\caption[*]{Integral contour of $X_{-}$ and branch cuts in three different
domains of $X_{+}$, where the straight line stands for the contour and the
photon line represents the branch cuts.}
\label{c1}
\end{figure}

\begin{figure}[tbp]
\begin{center}
\includegraphics[width=5cm]{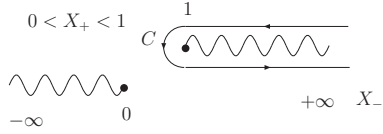}
\end{center}
\caption[*]{The contour $C$.}
\label{c2}
\end{figure}

\begin{eqnarray}
I &=&-\frac{i}{2}\int_{-\infty }^{+\infty }dX_{+}\int_{-\infty }^{+\infty
}dX_{-}\left[ X_{-}+i\epsilon \left( X_{+}-X_{-}\right) \right] ^{i-1}\left[
X_{+}-i\epsilon \left( X_{+}-X_{-}\right) \right] ^{\overline{i}-1}  \notag
\\
&&\times \left[ 1-X_{-}-i\epsilon \left( X_{+}-X_{-}\right) \right] ^{j-1}%
\left[ 1-X_{+}+i\epsilon \left( X_{+}-X_{-}\right) \right] ^{\overline{j}-1}.
\end{eqnarray}

The integral over $X_{+}$ can be decomposed into 3 pieces with respect to
three integration domains(see Fig.~\ref{c1}). Only in the second domain does
the integral give non-trivial contribution, since the contour of $X_{-}$ can
not be deformed to a single point only in that case. Therefore, we can reach
the factorized integrals,

\begin{equation}
I=\int_{0}^{1}dX_{+}X_{+}^{\overline{i}-1}\left( 1-X_{+}\right) ^{\overline{j%
}-1}\int_{C}\frac{dX_{-}}{2i}X_{-}^{i-1}\left( 1-X_{-}\right) ^{j-1},
\end{equation}
where contour $C$(see Fig.~\ref{c2}) is a contour which encloses the branch
cut $[1,\infty ]$. It starts from $i\epsilon +\infty $, goes to $i\epsilon
+1 $, then crosses the real axis to $-i\epsilon +1$, in the end, goes to $%
-i\epsilon +\infty $ and forms a close contour. It is straightforward to
compute this contour integral and obtain the final result,
\begin{eqnarray}
I &=&\int_{0}^{1}dX_{+}X_{+}^{\overline{i}-1}\left( 1-X_{+}\right) ^{%
\overline{j}-1}  \notag \\
&&\times \left[ \exp \left( -i\pi \left( j-1\right) \right) -\exp \left(
i\pi \left( j-1\right) \right) \right] \int_{1}^{\infty }\frac{dX_{-}}{2i}%
X_{-}^{i-1}\left( X_{-}-1\right) ^{j-1}, \\
&=&\pi \frac{\Gamma \left( \overline{i}\right) \Gamma \left( \overline{j}%
\right) }{\Gamma \left( \overline{i}+\overline{j}\right) }\frac{\Gamma
\left( 1-i-j\right) }{\Gamma \left( 1-i\right) \Gamma \left( 1-j\right) }%
=\pi \frac{\Gamma \left( i\right) \Gamma \left( j\right) }{\Gamma \left(
i+j\right) }\frac{\Gamma \left( 1-\overline{i}-\overline{j}\right) }{\Gamma
\left( 1-\overline{i}\right) \Gamma \left( 1-\overline{j}\right) }
\end{eqnarray}
Therefore, the $d^{2}w$ integral yields,
\begin{eqnarray*}
&&\dint d^{2}w\left[ \left( \frac{w-v}{2}\right) ^{-1+j}\left( \frac{w+v}{2}%
\right) ^{-1+k}\right] \left[ a.h.\right] \\
&=&4\pi \frac{\Gamma \left( k\right) \Gamma \left( j\right) }{\Gamma \left(
k+j\right) }\frac{\Gamma \left( 1-\overline{k}-\overline{j}\right) }{\Gamma
\left( 1-\overline{k}\right) \Gamma \left( 1-\overline{j}\right) }v^{j+k-1}%
\overline{v}^{\overline{j}+\overline{k}-1}.
\end{eqnarray*}

\subsection{The $d^{2}v$ integral}

The $d^{2}v$ integral can be cast into

\begin{eqnarray}
&&\dint d^{2}v\exp \left( iq_{-}\cdot v\right) v^{\eta -2}\overline{v}^{%
\overline{\eta }-2}, \\
&=&\int_{0}^{\infty }dr\int_{0}^{2\pi }d\varphi r^{\eta +\overline{\eta }%
-3}\exp \left[ i\left( \eta -\overline{\eta }\right) \varphi +iq_{-}r\cos
\varphi \right] , \\
&=&i^{\eta -\overline{\eta }}\pi \left( \frac{2}{q_{-}}\right) ^{\eta +%
\overline{\eta }-2}\frac{\Gamma \left( \eta -1\right) }{\Gamma \left( 2-%
\overline{\eta }\right) },
\end{eqnarray}
where $\eta =i+j+k$ and $\eta -\overline{\eta }=-\left(
n+n_{a}+n_{b}\,\right) \in Z$. In reaching the above result, we have set the
orientation of the $\overrightarrow{q}_{-}$ parallel to the real axis$\left(
\varphi _{q_{-}}=0\right) $, and used the following two formulae:%
\begin{eqnarray}
J_{m}\left( z\right) &=&\frac{1}{2\pi i^{m}}\int_{0}^{2\pi }\exp \left(
iz\cos \varphi +im\varphi \right) , \\
\int_{0}^{\infty }dxx^{\mu }J_{\nu }\left( ax\right) &=&\frac{2^{\mu }}{%
a^{\mu +1}}\frac{\Gamma \left( \frac{1}{2}+\frac{\nu }{2}+\frac{\mu }{2}%
\right) }{\Gamma \left( \frac{1}{2}+\frac{\nu }{2}-\frac{\mu }{2}\right) },
\end{eqnarray}
where $m$ should be integers.

Combining the two results, one gets,
\begin{equation}
I_{Q,q_{-}}^{h,h_{a},h_{b},0}=4\pi ^{2}i^{-\left( n+n_{a}+n_{b}\,\right)
}\left( \frac{4}{q_{-}^{2}}\right) ^{\gamma +\gamma _{a}+\gamma _{b}-1}\frac{%
\Gamma \left( h+h_{a}-h_{b}\right) \Gamma \left( h+h_{b}-h_{a}\right) \Gamma
\left( 1-2\overline{h}\right) }{\Gamma \left( 2h\right) \Gamma \left( 1-%
\overline{h}-\overline{h}_{a}+\overline{h}_{b}\right) \Gamma \left( 1-%
\overline{h}-\overline{h}_{b}+\overline{h}_{a}\right) }\frac{\Gamma \left(
h+h_{a}+h_{b}-1\right) }{\Gamma \left( 2-\overline{h}-\overline{h}_{a}-%
\overline{h}_{b}\right) }.
\end{equation}

\section{Other terms in the Taylor expansion}

\label{general}

In this part, we discuss cases of the higher terms of the Taylor series of $%
\exp \left( iQ\frac{w}{2}\right) $.

\subsection{The second term in the expansion}

The relevant integral of the second term in the expansion reads,
\begin{eqnarray}
I_{Q,q_{-}}^{h,h_{a},h_{b},1} &=&\dint d^{2}vd^{2}w\exp \left( iq_{-}\cdot
v\right) \left[ v^{-1+i}\left( \frac{w-v}{2}\right) ^{-1+j}\left( \frac{w+v}{%
2}\right) ^{-1+k}\right] \left[ a.h.\right] i\frac{Q}{2}\cdot \left[ \left(
\frac{w-v}{2}\right) +\left( \frac{w+v}{2}\right) \right] \\
&=&\dint d^{2}v\exp \left( iq_{-}\cdot v\right) \left( i\frac{Q}{2}\cdot
v\right) v^{\eta -2}\overline{v}^{\overline{\eta }-2}4\pi \frac{\Gamma
\left( k\right) \Gamma \left( j\right) }{\Gamma \left( 1+k+j\right) }\frac{%
\Gamma \left( 1-\overline{k}-\overline{j}\right) }{\Gamma \left( 1-\overline{%
k}\right) \Gamma \left( 1-\overline{j}\right) }.
\end{eqnarray}%
where $I_{Q,q_{-}}^{h,h_{a},h_{b},l}$ stands for the relevant integral of
the $l+1$ term in the Taylor expansion. Assume $i\frac{Q}{2}\cdot v=i\frac{Q%
}{2}v\cos \left( \phi +\varphi \right) $, where $\phi $ is the orientation
of $\overrightarrow{Q}$ in 2-dim plane. $\phi $ should also be considered as
the angle between $\overrightarrow{q}_{-}$ and $\overrightarrow{Q}$. It is
straightforward to compute the $v$ integral and obtain
\begin{eqnarray}
I_{Q,q_{-}}^{h,h_{a},h_{b},1} &=&4\pi ^{2}i^{-\left( n+n_{a}+n_{b}\,\right) }%
\frac{Q}{2q_{-}}\left( \frac{4}{q_{-}^{2}}\right) ^{\gamma +\gamma
_{a}+\gamma _{b}-1}\frac{\Gamma \left( k\right) \Gamma \left( j\right) }{%
\Gamma \left( 1+k+j\right) }\frac{\Gamma \left( 1-\overline{k}-\overline{j}%
\right) }{\Gamma \left( 1-\overline{k}\right) \Gamma \left( 1-\overline{j}%
\right) }\left( k-j\right)  \notag \\
&&\times \left[ \exp \left( -i\phi \right) \frac{\Gamma \left(
i+j+k-1\right) }{\Gamma \left( 1-\overline{i}-\overline{j}-\overline{k}%
\right) }-\exp \left( i\phi \right) \frac{\Gamma \left( i+j+k-1\right) }{%
\Gamma \left( 1-\overline{i}-\overline{j}-\overline{k}\right) }\right] .
\end{eqnarray}%
Since $k-j=2\left( h_{a}-h_{b}\right) $ and eventually $h_{a}=h_{b}=\frac{1}{%
2}$, $I_{Q,q_{-}}^{h,h_{a},h_{b},1}$ vanishes in the end. Moreover, it is
easy to prove that integrals of all odd power of $Q\frac{w}{2}$ are
proportional to $k-j$ and they vanish as well.

\subsection{The third term in the expansion}

The relevant integral of the third term in the expansion reads,

\begin{eqnarray}
I_{Q,q_{-}}^{h,h_{a},h_{b},2} &=&\dint d^{2}vd^{2}w\exp \left( iq_{-}\cdot
v\right) \left[ v^{-1+i}\left( \frac{w-v}{2}\right) ^{-1+j}\left( \frac{w+v}{%
2}\right) ^{-1+k}\right] \left[ a.h.\right] \frac{1}{2}\left( i\frac{Q}{2}%
\cdot w\right) ^{2}, \\
&=&\dint d^{2}v\exp \left( iq_{-}\cdot v\right) \frac{1}{2}\left( i\frac{Q}{2%
}\cdot v\right) ^{2}v^{\eta -2}\overline{v}^{\overline{\eta }-2}  \notag \\
&&\times 4\pi \frac{\Gamma \left( k\right) \Gamma \left( j\right) }{\Gamma
\left( 2+k+j\right) }\frac{\Gamma \left( 1-\overline{k}-\overline{j}\right)
}{\Gamma \left( 1-\overline{k}\right) \Gamma \left( 1-\overline{j}\right) }%
\left[ \left( k-j\right) ^{2}+\left( k+j\right) \right] .
\end{eqnarray}%
Dropping the $\left( k-j\right) ^{2}$ term and finishing the $\dint d^{2}v$
integral, one gets%
\begin{eqnarray}
I_{Q,q_{-}}^{h,h_{a},h_{b},2} &=&-4\pi ^{2}i^{-\left( n+n_{a}+n_{b}\,\right)
}\frac{Q^{2}}{4q_{-}^{2}}\left( \frac{4}{q_{-}^{2}}\right) ^{\gamma +\gamma
_{a}+\gamma _{b}-1}\frac{\Gamma \left( k\right) \Gamma \left( j\right) }{%
\Gamma \left( 2+k+j\right) }\frac{\Gamma \left( 1-\overline{k}-\overline{j}%
\right) }{\Gamma \left( 1-\overline{k}\right) \Gamma \left( 1-\overline{j}%
\right) }\left( k+j\right)  \notag \\
&&\times \left[ \frac{\Gamma \left( \eta \right) }{\Gamma \left( 1-\overline{%
\eta }\right) }-\frac{\exp \left( 2i\phi \right) }{2}\frac{\Gamma \left(
\eta +1\right) }{\Gamma \left( 2-\overline{\eta }\right) }-\frac{\exp \left(
-2i\phi \right) }{2}\frac{\Gamma \left( \eta -1\right) }{\Gamma \left( -%
\overline{\eta }\right) }\right] .
\end{eqnarray}%
Thus, it is straightforward to examine the singularity structure of $%
n_{Y,Q,q_{-},2}^{(2)\rho ,\rho _{a},\rho _{b}}$(it is proportional to $%
I_{Q,q_{-}}^{h,h_{a},h_{b},2}$ which comes from the third term in the
expansion) before the final $\gamma $ integral in the complex $\gamma $
plane when $n=0$.
\begin{eqnarray}
n_{Y,Q,q_{-},2}^{(2)\rho ,\rho _{a},\rho _{b}} &\propto &\int \frac{d\gamma
}{2\pi i}2\pi \frac{2\nu ^{2}}{\pi ^{4}}\frac{\left\vert \rho \right\vert }{%
2\chi \left( \frac{1}{2}\right) -\chi \left( \gamma \right) }E_{Q}^{0,\nu
}\left( \rho \right) g_{3P}\left( 1-\gamma ,\frac{1}{2},\frac{1}{2}\right)
I_{Q,q_{-}}^{h,h_{a},h_{b},2}  \notag \\
&\propto &\int \frac{d\gamma }{2\pi i}\frac{\left\vert \rho \right\vert }{%
2\chi \left( \frac{1}{2}\right) -\chi \left( \gamma \right) }E_{Q}^{0,\nu
}\left( \rho \right) g_{3P}\left( 1-\gamma ,\frac{1}{2},\frac{1}{2}\right)
\frac{1}{\gamma +\frac{1}{2}}\left[ \frac{\Gamma \left( \gamma +1\right) }{%
\Gamma \left( -\gamma \right) }-\cos 2\phi \frac{\Gamma \left( \gamma
+2\right) }{\Gamma \left( 1-\gamma \right) }\right] .
\end{eqnarray}%
In addition, one can also show that the fifth term in the expansion
\begin{eqnarray}
n_{Y,Q,q_{-},4}^{(2)\rho ,\rho _{a},\rho _{b}} &\propto &\int \frac{d\gamma
}{2\pi i}2\pi \frac{2\nu ^{2}}{\pi ^{4}}\frac{\left\vert \rho \right\vert }{%
2\chi \left( \frac{1}{2}\right) -\chi \left( \gamma \right) }E_{Q}^{0,\nu
}\left( \rho \right) g_{3P}\left( 1-\gamma ,\frac{1}{2},\frac{1}{2}\right)
I_{Q,q_{-}}^{h,h_{a},h_{b},4}  \notag \\
&\propto &\int \frac{d\gamma }{2\pi i}\frac{\left\vert \rho \right\vert }{%
2\chi \left( \frac{1}{2}\right) -\chi \left( \gamma \right) }E_{Q}^{0,\nu
}\left( \rho \right) g_{3P}\left( 1-\gamma ,\frac{1}{2},\frac{1}{2}\right)
\frac{1}{\left( \gamma +\frac{1}{2}\right) \left( \gamma +\frac{3}{2}\right)
}f_{4}\left[ \gamma ,\phi \right] ,
\end{eqnarray}
where $f_{4}\left[ \gamma ,\phi \right] $ are terms which are singular only
at integer values of $\gamma $. Therefore, we can see the pattern and
conclude that $\gamma =0.18$ and $0.82$ are the only two singularities in
the $\left( 0,1\right) $ domain when $n=0$. In the case when $n\neq 0$,
since things are much more complicated, we have been unable to reach similar
conclusion. Therefore, we have to restrict ourselves to the forward
scattering case in which $Q=0$. Nevertheless, contributions from
sub-dominant trajectories would not play any role when one averages over the
orientation of the initial onium, since those contributions are all angular
dependent and vanish after averaging over angles.

\section{Saddle point evaluation of the pomeron}

\label{saddle}

In this appendix, we use saddle point approximation to evaluate the pomeron
trajectory when $Y$ is large. The relevant integral is
\begin{eqnarray}
&&\dint dh_{a}\frac{b_{n_{a},\nu _{a}}\left\vert \rho _{a}\right\vert }{2\pi
^{2}}E_{q_{a}}^{n_{a},\nu _{a}}\left( \rho _{a}\right) \left( \frac{4}{q^{2}}%
\right) ^{\gamma _{a}}\exp \left[ \omega \left( a\right) Y\right] , \\
&\simeq &\dint_{-\infty }^{+\infty }\frac{d\nu _{a}}{\pi ^{4}}2\nu
_{a}^{2}\int d^{2}R\exp \left( iq_{a}\cdot R\right) \left( \frac{2\rho _{a}}{%
\left\vert R-\frac{\rho _{a}}{2}\right\vert \left\vert R+\frac{\rho _{a}}{2}%
\right\vert q}\right) ^{1+2i\nu _{a}}\exp \left[ \omega \left( a\right) Y%
\right] , \\
&\simeq &\int d^{2}R\frac{\exp \left( iq_{a}\cdot R\right) 2\rho _{a}}{%
\left\vert R-\frac{\rho _{a}}{2}\right\vert \left\vert R+\frac{\rho _{a}}{2}%
\right\vert q}\dint_{-\infty }^{+\infty }\frac{d\nu _{a}}{\pi ^{4}}2\nu
_{a}^{2}\exp \left[ 2\ln 2\overline{\alpha }_{s}Y-7\xi \left( 3\right) \nu
_{a}^{2}\overline{\alpha }_{s}Y+2i\nu _{a}\ln \frac{2\rho _{a}}{\left\vert R-%
\frac{\rho _{a}}{2}\right\vert \left\vert R+\frac{\rho _{a}}{2}\right\vert q}%
\right] ,
\end{eqnarray}%
where $\overline{\alpha }_{s}=\frac{4\alpha _{s}C_{F}}{\pi }$. In reaching
the above result, we have used the idea of saddle point approximation and
neglected all sub-dominant trajectories. In the region where the diffusion
approximation $\ln ^{2}\left( q\rho \right) \ll \frac{14\alpha C_{F}\zeta
\left( 3\right) }{\pi }Y$ is valid, this integral yields%
\begin{equation}
\frac{1}{2\pi }\frac{\rho _{a}}{q}e\left( \rho _{a},q_{a}\right) \frac{\exp %
\left[ \left( \alpha _{P}-1\right) Y\right] \exp \left[ -\frac{\pi \ln
^{2}\left( \rho _{a}q_{a}\right) }{28\alpha C_{F}\zeta \left( 3\right) Y}%
\right] }{\left( 7\alpha \zeta \left( 3\right) C_{F}Y\right) ^{3/2}},
\end{equation}%
where we have set $\ln ^{2}\frac{2\rho _{a}}{\left\vert R-\frac{\rho _{a}}{2}%
\right\vert \left\vert R+\frac{\rho _{a}}{2}\right\vert q}\simeq \ln
^{2}\left( \rho _{a}q_{a}\right) $ since $\int d^{2}R$ is dominated in the
region where $R\sim 1/q_{a}$, and we have defined $e\left( \rho
_{a},q_{a}\right) =\frac{1}{2\pi }\int d^{2}R\exp \left( iq_{a}R\right)
\frac{1}{\left\vert R-\frac{\rho _{a}}{2}\right\vert \left\vert R+\frac{\rho
_{a}}{2}\right\vert }$, $\alpha _{P}-1=\frac{8\ln 2\alpha _{s}C_{F}}{\pi }$.

\section{Explicit calculation of $V_{\protect\nu }$}

\label{v}

According to Eq.(\ref{mellin}) and Mellin transform, one can easily obtain,
\begin{equation}
\int_{\frac{1}{2}-i\infty }^{\frac{1}{2}+i\infty }\frac{d\gamma }{2\pi i}%
\left( q^{2}x_{01}^{2}\right) ^{\gamma -1}V_{\gamma }=\int \frac{d^{2}x_{2}}{%
x_{01}x_{12}}J_{0}\left( \frac{1}{2}qx_{01}\right) e\left( x_{12},q\right)
e\left( x_{02},q\right) .
\end{equation}%
Evaluating the above integral in $q^{2}x_{01}^{2}\ll 1$ limit, we should
close the contour to the right half plane. The leading order term tells us
the position of the first pole of $V_{\gamma }$ in complex $\gamma $ plane.

First of all, it is straight forward to discover that
\begin{equation}
e\left( x,q\right) =\frac{1}{\pi }\int_{0}^{1}du\frac{\exp \left[ iq\cdot
x\left( \frac{1}{2}-u\right) \right] }{\left[ u\left( 1-u\right) \right]
^{1/2}}K_{0}\left[ qx\sqrt{u\left( 1-u\right) }\right] ,
\end{equation}%
where we have used the formulae
\begin{equation}
\frac{1}{\sqrt{ab}}=\frac{1}{\Gamma ^{2}\left( \frac{1}{2}\right) }%
\int_{0}^{1}du\frac{1}{\left[ u\left( 1-u\right) \right] ^{1/2}}\frac{1}{%
au+b\left( 1-u\right) },
\end{equation}
and
\begin{equation}
\int_{0}^{\infty }\frac{J_{\nu }\left( qx\right) x^{\nu +1}}{\left(
x^{2}+a^{2}\right) ^{\mu +1}}dx=\frac{a^{\nu -\mu }q^{\mu }K_{\nu -\mu
}\left( qa\right) }{2^{\mu }\Gamma \left( \mu +1\right) }.
\end{equation}
Changing variable $u=\sin ^{2}\frac{t}{2}$, and combining
\begin{eqnarray}
K_{\mu }\left( x\right) &=&\frac{\pi }{2\sin \pi \mu }\left[ I_{-\mu }\left(
x\right) -I_{\mu }\left( x\right) \right] , \\
I_{\mu }\left( x\right) &=&\exp \left( -i\frac{\pi }{2}\mu \right) J_{\mu }%
\left[ \exp \left( -i\frac{\pi }{2}\right) x\right]
\end{eqnarray}%
along with the integration identity
\begin{equation}
\int_{0}^{\pi /2}J_{\mu }\left[ z\sin t\right] \cos \left( x\cos t\right) dt=%
\frac{\pi }{2}J_{\frac{\mu }{2}}\left[ y_{+}\right] J_{\frac{\mu }{2}}\left[
y_{-}\right] ,
\end{equation}%
where $y_{\pm }=\frac{\sqrt{x^{2}+z^{2}}\pm x}{2}$, gives
\begin{eqnarray}
e\left( x,q\right) &=&\lim_{\mu \rightarrow 0}\frac{\pi }{2\sin \pi \mu }%
\left[ J_{-\frac{\mu }{2}}\left( \varrho \right) J_{-\frac{\mu }{2}}\left(
\varrho ^{\ast }\right) -J_{\frac{\mu }{2}}\left( \varrho \right) J_{\frac{%
\mu }{2}}\left( \varrho ^{\ast }\right) \right] , \\
&=&-\frac{\pi }{4}\left[ Y_{0}\left( \varrho \right) J_{0}\left( \varrho
^{\ast }\right) +J_{0}\left( \varrho \right) Y_{0}\left( \varrho ^{\ast
}\right) \right] ,
\end{eqnarray}%
where $\varrho =\frac{qx}{4}e^{i\psi }$ with $\psi $ being the angle between
$\overrightarrow{q}$ and $\overrightarrow{x}$. Therefore,
\begin{eqnarray}
&&\int \frac{d^{2}x_{2}}{x_{01}x_{12}}J_{0}\left( \frac{1}{2}qx_{01}\right)
e\left( x_{12},q\right) e\left( x_{02},q\right)  \notag \\
&=&\frac{\pi ^{3}}{8}J_{0}\left( \frac{qx_{10}}{2}\right) \int
dx_{12}dx_{20}dbbJ_{0}\left( bx_{01}\right) J_{0}\left( bx_{02}\right)
J_{0}\left( bx_{12}\right)  \notag \\
&&\times \left[ Y_{0}\left( \varrho _{12}\right) J_{0}\left( \varrho
_{12}^{\ast }\right) +J_{0}\left( \varrho _{12}\right) Y_{0}\left( \varrho
_{12}^{\ast }\right) \right] \left[ Y_{0}\left( \varrho _{20}\right)
J_{0}\left( \varrho _{20}^{\ast }\right) +J_{0}\left( \varrho _{20}\right)
Y_{0}\left( \varrho _{20}^{\ast }\right) \right] ,
\end{eqnarray}%
where the identities\cite{Mueller:1993rr}
\begin{equation}
d^{2}x_{2}=J\left( x_{21},x_{20}\right) dx_{21}dx_{20}=\frac{4x_{21}x_{20}}{%
\sqrt{\left[ \left( x_{21}+x_{20}\right) ^{2}-x_{10}^{2}\right] \left[
x_{10}^{2}-\left( x_{21}-x_{20}\right) ^{2}\right] }},
\end{equation}%
and
\begin{equation}
\frac{\pi }{2}\int dbbJ_{0}\left( bx_{01}\right) J_{0}\left( bx_{02}\right)
J_{0}\left( bx_{12}\right) =\frac{1}{\sqrt{\left[ \left(
x_{21}+x_{20}\right) ^{2}-x_{10}^{2}\right] \left[ x_{10}^{2}-\left(
x_{21}-x_{20}\right) ^{2}\right] }}
\end{equation}%
have been used. It is easy to estimate the above integral and obtain $\int
\frac{d^{2}x_{2}}{x_{01}x_{12}}J_{0}\left( \frac{1}{2}qx_{01}\right) e\left(
x_{12},q\right) e\left( x_{02},q\right) \sim \ln \frac{1}{q^{2}x_{10}^{2}}$,
which indicates that the first pole of $V_{\gamma }$ occurs at $\gamma =1$.

\section{Detailed calculation for higher pomeron trajectories.}

\label{highn}

Here we consider the integrand of $\int d\gamma $ in the case of general
value of $n$,
\begin{eqnarray}
Integrand &=&2\pi \frac{2\left( \nu ^{2}+n^{4}/4\right) }{\pi ^{4}}b_{n,\nu
}I_{Q,q_{-}}^{h,h_{a},h_{b},0} \\
&=&\frac{4\left( \nu ^{2}+n^{4}/4\right) }{\pi ^{3}}\frac{\pi ^{3}2^{4i\nu }%
}{\left\vert n\right\vert /2-i\nu }\frac{\Gamma \left( \left\vert
n\right\vert /2-i\nu +1/2\right) \Gamma \left( \left\vert n\right\vert
/2+i\nu \right) }{\Gamma \left( \left\vert n\right\vert /2+i\nu +1/2\right)
\Gamma \left( \left\vert n\right\vert /2-i\nu \right) }  \notag \\
&&\times 4\pi ^{2}i^{-n}\left( \frac{4}{q^{2}}\right) ^{\gamma }\frac{\Gamma
^{3}\left( h\right) \Gamma \left( 1-2\overline{h}\right) }{\Gamma \left(
2h\right) \Gamma ^{3}\left( 1-\overline{h}\right) }
\end{eqnarray}%
When $n>0$, the integrand yields $\left( -i\right) ^{-\left\vert
n\right\vert }8\pi ^{2}\left( \frac{4}{q^{2}}\right) ^{\gamma }\frac{\Gamma
\left( \gamma +\frac{\left\vert n\right\vert }{2}\right) }{\Gamma \left( 1+%
\frac{\left\vert n\right\vert }{2}-\gamma \right) }$; when $n<0$, it gives $%
i^{\left\vert n\right\vert }8\pi ^{2}\left( \frac{4}{q^{2}}\right) ^{\gamma }%
\frac{\Gamma \left( \gamma +\frac{\left\vert n\right\vert }{2}\right) }{%
\Gamma \left( 1+\frac{\left\vert n\right\vert }{2}-\gamma \right) }$.
Generally, it can be cast into $i^{\left\vert n\right\vert }8\pi ^{2}\left(
\frac{4}{q^{2}}\right) ^{\gamma }\frac{\Gamma \left( \gamma +\frac{%
\left\vert n\right\vert }{2}\right) }{\Gamma \left( 1+\frac{\left\vert
n\right\vert }{2}-\gamma \right) }$.

\section{Triple pomeron coefficient $g_{3P}$}

\label{g3p}

In this appendix, following Korchemsky\cite{Korchemsky:1997fy}, we discuss
the triple pomeron coefficient $g_{3P}$.
\begin{eqnarray}
g_{3P}\left( h,h_{a},h_{b}\right)  &=&\dint \frac{d^{2}\rho _{2}d^{2}\rho
_{3}d^{2}\rho _{4}}{\left\vert \rho _{23}\rho _{34}\rho _{42}\right\vert ^{2}%
}\rho _{23}^{h}\overline{\rho }_{23}^{\overline{h}}\left( \frac{\rho _{24}}{%
\rho _{2}\rho _{4}}\right) ^{h_{a}}\left( \frac{\overline{\rho }_{24}}{%
\overline{\rho }_{2}\overline{\rho }_{4}}\right) ^{\overline{h}_{a}}  \notag
\\
&&\times \left( \frac{\rho _{34}}{\left( 1-\rho _{3}\right) \left( 1-\rho
_{4}\right) }\right) ^{h_{b}}\left( \frac{\overline{\rho }_{34}}{\left( 1-%
\overline{\rho }_{3}\right) \left( 1-\overline{\rho }_{4}\right) }\right) ^{%
\overline{h}_{b}}
\end{eqnarray}%
Changing the variables: $\rho ^{\prime }=\overline{\rho }$ and $\overline{%
\rho }^{\prime }=\rho $,
\begin{eqnarray}
g_{3P}\left( h,h_{a},h_{b}\right)  &=&\dint \frac{d^{2}\rho _{2}^{\prime
}d^{2}\rho _{3}^{\prime }d^{2}\rho _{4}^{\prime }}{\left\vert \rho
_{23}^{\prime }\rho _{34}^{\prime }\rho _{42}^{\prime }\right\vert ^{2}}%
\overline{\rho }_{23}^{\prime h}\rho _{23}^{\prime \overline{h}}\left( \frac{%
\overline{\rho }_{24}^{\prime }}{\overline{\rho }_{2}^{\prime }\overline{%
\rho }_{4}}\right) ^{h_{a}}\left( \frac{\rho _{24}^{\prime }}{\rho
_{2}^{\prime }\rho _{4}^{\prime }}\right) ^{\overline{h}_{a}}  \notag \\
&&\times \left( \frac{\overline{\rho }_{34}^{\prime }}{\left( 1-\overline{%
\rho }_{3}^{\prime }\right) \left( 1-\overline{\rho }_{4}^{\prime }\right) }%
\right) ^{h_{b}}\left( \frac{\rho _{34}^{\prime }}{\left( 1-\rho
_{3}^{\prime }\right) \left( 1-\rho _{4}^{\prime }\right) }\right) ^{%
\overline{h}_{b}} \\
&=&g_{3P}\left( \overline{h},\overline{h}_{a},\overline{h}_{b}\right) ,
\end{eqnarray}%
which means triple pomeron coefficient $g_{3P}$ is an even function of $n$, $%
n_{a}$ and $n_{b}$. Following the methods used in ref.\cite%
{Korchemsky:1997fy}(here our $h$ corresponds the $h_{\gamma }$ in ref.\cite%
{Korchemsky:1997fy}), we find that
\begin{equation}
g_{3P}\left( h,\frac{1}{2},\frac{1}{2}\right) =\frac{1}{\Gamma \left(
1-h\right) }\sum_{c=1}^{3}J_{c}\left( h,\frac{1}{2},\frac{1}{2}\right)
\overline{J}_{c}\left( \overline{h},\frac{1}{2},\frac{1}{2}\right)
\end{equation}%
where $J_{c}\left( h,\frac{1}{2},\frac{1}{2}\right) $ and $\overline{J}%
_{c}\left( \overline{h},\frac{1}{2},\frac{1}{2}\right) $ can be written in
terms of Meijer' G functions $G_{44}^{24}$ and hypergeometric functions $%
\left. _{4}F_{3}\right. $ as follows
\begin{equation}
J_{1}\left( h,\frac{1}{2},\frac{1}{2}\right) =\Gamma \left( 1-h\right)
G_{44}^{42}\left( 1\QATOPD\vert . {1,\frac{1}{2}+h,1,\frac{3}{2}-h}{\frac{1}{%
2},\frac{1}{2},\frac{1}{2},\frac{1}{2}}\right) =\Gamma \left( 1-h\right)
G_{44}^{24}\left( 1\QATOPD\vert . {\frac{1}{2},\frac{1}{2},\frac{1}{2},\frac{%
1}{2}}{0,\frac{1}{2}-h,0,h-\frac{1}{2}}\right) ;
\end{equation}%
\begin{equation}
J_{2}\left( h,\frac{1}{2},\frac{1}{2}\right) =J_{3}\left( h,\frac{1}{2},%
\frac{1}{2}\right) =\frac{\pi ^{2}\Gamma ^{3}\left( 1-h\right) }{\Gamma
\left( \frac{3}{2}-h\right) \Gamma \left( \frac{3}{2}-h\right) }\left.
_{4}F_{3}\right. \left( \QATOPD. \vert {\frac{1}{2},\frac{1}{2},1-h,1-h}{1,%
\frac{3}{2}-h,\frac{3}{2}-h}1\right) ;
\end{equation}%
\begin{equation}
J_{3}\left( h,h_{a},h_{b}\right) =J_{2}\left( h,h_{b},h_{a}\right) ;
\end{equation}%
\begin{equation}
\overline{J}_{1}\left( \overline{h},\frac{1}{2},\frac{1}{2}\right) =\frac{%
\pi 2^{-\overline{h}}}{\Gamma \left( 1-\overline{h}\right) \Gamma \left(
\overline{h}\right) }G_{44}^{42}\left( 1\QATOPD\vert . {1,\frac{3}{2}-%
\overline{h},1,1}{\frac{1}{2},\frac{1}{2},\frac{1}{2}-\frac{1}{2}\overline{h}%
,1-\frac{1}{2}\overline{h}}\right) =\frac{\pi 2^{-\overline{h}}}{\Gamma
\left( 1-\overline{h}\right) \Gamma \left( \overline{h}\right) }%
G_{44}^{24}\left( 1\QATOPD\vert . {\frac{1}{2},\frac{1}{2},\frac{1}{2}%
\overline{h}+\frac{1}{2},\frac{1}{2}\overline{h}}{0,\overline{h}-\frac{1}{2}%
,0,0}\right) ;
\end{equation}%
\begin{equation}
\overline{J}_{2}\left( \overline{h},\frac{1}{2},\frac{1}{2}\right) =%
\overline{J}_{3}\left( \overline{h},\frac{1}{2},\frac{1}{2}\right)
=G_{44}^{42}\left( 1\QATOPD\vert . {1,\frac{3}{2}-\overline{h},1,\frac{1}{2}+%
\overline{h}}{\frac{1}{2},\frac{1}{2},\frac{1}{2},\frac{1}{2}}\right)
=G_{44}^{24}\left( 1\QATOPD\vert . {\frac{1}{2},\frac{1}{2},\frac{1}{2},%
\frac{1}{2}}{0,\overline{h}-\frac{1}{2},0,\frac{1}{2}-\overline{h}}\right) ;
\end{equation}%
\begin{equation}
\overline{J}_{3}\left( \overline{h},\overline{h}_{a},\overline{h}_{b}\right)
=\overline{J}_{2}\left( \overline{h},\overline{h}_{b},\overline{h}%
_{a}\right) .
\end{equation}%
We find different but equivalent final expressions of $J_{c}\left(
h,h_{a},h_{b}\right) $
and $\overline{J}_{c}\left( \overline{h},\overline{h}_{a},\overline{h}%
_{b}\right) $ as compared to those in
ref.\cite{Korchemsky:1997fy}. \footnote{Due to some numerical
subtleties, B.X. did not realize the equivalence between the
results obtained above and the final expressions of $J_{c}\left(
h,h_{a},h_{b}\right) $
and $\overline{J}_{c}\left( \overline{h},\overline{h}_{a},\overline{h}%
_{b}\right)$ in ref.\cite{Korchemsky:1997fy}. B.X. would like to
thank Dr. L. Motyka for communications on this issue.}
Numerically, one finds $g_{3P}\left( \frac{1}{2},\frac{1}{2},\frac{1}{2}%
\right) =7766.68$ which agrees with the previous result in ref.\cite%
{Korchemsky:1997fy,Bialas:1997ig}. Thus, we outline our evaluation of the $%
g_{3P}\left( h,\frac{1}{2},\frac{1}{2}\right) $ in the following. For
instance, from ref.\cite{Korchemsky:1997fy}, one can get
\begin{equation}
\overline{J}_{1}\left( \overline{h},\frac{1}{2},\frac{1}{2}\right) =\frac{%
\pi ^{3}}{\Gamma \left( 1-\overline{h}\right) \Gamma \left( \overline{h}%
\right) }\int_{0}^{1}dxx^{-\overline{h}}\left( 1-x\right) ^{-1+\overline{h}%
}\left. _{2}F_{1}\right. \left( \QATOPD. \vert {\frac{1}{2},\frac{1}{2}%
}{1}x\right) \left. _{2}F_{1}\right. \left( \QATOPD. \vert {\frac{1}{2},%
\frac{1}{2}}{1}x\right) .
\end{equation}%
Using the identity\cite{Bialas:1997ig}
\begin{equation}
\left. _{2}F_{1}\right. \left( \QATOPD. \vert {\frac{1}{2},\frac{1}{2}%
}{1}x\right) \left. _{2}F_{1}\right. \left( \QATOPD. \vert {\frac{1}{2},%
\frac{1}{2}}{1}x\right) =\frac{1}{\left( 1-x\right) ^{1/2}}\left.
_{3}F_{2}\right. \left( \QATOPD. \vert {\frac{1}{2},\frac{1}{2},\frac{1}{2}%
}{1,1}-\frac{x^{2}}{4\left( 1-x\right) }\right) ,
\end{equation}%
and the Mellin-Barnes representation of hypergeometric function
\begin{equation}
\left. _{3}F_{2}\right. \left( \QATOPD. \vert {\frac{1}{2},\frac{1}{2},\frac{%
1}{2}}{1,1}z\right) =\frac{1}{\Gamma ^{3}\left( \frac{1}{2}\right) }%
\int_{-i\infty }^{+i\infty }\frac{ds}{2\pi i}\frac{\Gamma ^{3}\left( \frac{1%
}{2}+s\right) \Gamma \left( -s\right) }{\Gamma ^{2}\left( 1+s\right) }\left(
-z\right) ^{s},
\end{equation}%
along with identities
\begin{eqnarray}
\int_{0}^{1}dxx^{\mu -1}\left( 1-x\right) ^{\nu -1} &=&\frac{\Gamma \left(
\mu \right) \Gamma \left( \nu \right) }{\Gamma \left( \mu +\nu \right) }, \\
\Gamma \left( 2s\right)  &=&\frac{2^{2s-1}}{\sqrt{\pi }}\Gamma \left(
s\right) \Gamma \left( s+\frac{1}{2}\right) ,
\end{eqnarray}%
one obtains,
\begin{eqnarray}
\overline{J}_{1}\left( \overline{h},\frac{1}{2},\frac{1}{2}\right)  &=&\frac{%
\pi 2^{-\overline{h}}}{\Gamma \left( 1-\overline{h}\right) \Gamma \left(
\overline{h}\right) }  \notag \\
&&\times \int_{-i\infty }^{+i\infty }\frac{ds}{2\pi i}\frac{\Gamma
^{2}\left( \frac{1}{2}+s\right) \Gamma \left( s+\frac{1}{2}-\frac{1}{2}%
\overline{h}\right) \Gamma \left( s+1-\frac{1}{2}\overline{h}\right) \Gamma
\left( -s\right) \Gamma \left( \overline{h}-\frac{1}{2}-s\right) }{\Gamma
^{2}\left( 1+s\right) }.
\end{eqnarray}%
According to the definition of the Meijer's G function%
\begin{equation}
G_{pq}^{mn}\left( z\QATOPD\vert . {a_{1},\ldots ,a_{p}}{b_{1},\ldots
,b_{q}}\right) =\int_{-i\infty }^{+i\infty }\frac{ds}{2\pi i}\frac{%
\dprod\limits_{j=1}^{m}\Gamma \left( b_{j}+s\right)
\dprod\limits_{j=1}^{n}\Gamma \left( 1-a_{j}-s\right) }{\dprod%
\limits_{j=m+1}^{q}\Gamma \left( 1-b_{j}-s\right)
\dprod\limits_{j=n+1}^{p}\Gamma \left( a_{j}+s\right) }z^{-s},
\end{equation}%
one can easily reach the expression we found above. It is straightforward to
put $g_{3P}\left( h,\frac{1}{2},\frac{1}{2}\right) $ into Mathematica and
plot it for fixed integer values of $n$ as function of $\gamma $. The plots
for the first few integer values of $n$ show that $g_{3P}\left( h,\frac{1}{2}%
,\frac{1}{2}\right) $ has no pole in the domain $\left( -\frac{\left\vert
n\right\vert }{2},\frac{\left\vert n\right\vert }{2}+1\right) $.


\begin{thebibliography}{99}
\bibitem{Munier} Private communication with Stephane Munier. I would like to
thank Stephane Munier for the notes about the computation of some integrals.
\end{thebibliography}
\end{document}